%% file: MonoZ.tex
\documentclass[11pt,a4paper]{article}
\pdfoutput=1
\usepackage{jheppub}

\usepackage{xspace}
\usepackage{epsfig}
\usepackage{amssymb,amsmath}
\usepackage{url,booktabs}
\usepackage[usenames,table,dvipsnames]{xcolor}
\usepackage{slashed}
\usepackage{subfig}
\usepackage{multirow}
\usepackage{tikz}
\usepackage{bm}
\usepackage[utf8]{inputenc}
\usetikzlibrary{arrows,shapes}
\usetikzlibrary{decorations.markings}
\usetikzlibrary{matrix,arrows} 	
\usetikzlibrary{trees}			
\usetikzlibrary{positioning}	
\usetikzlibrary{calc,through}	
\usetikzlibrary{angles,quotes}
\usetikzlibrary{decorations.pathreplacing}  
\usepackage{pgffor}
\usetikzlibrary{decorations.pathmorphing}	
\usetikzlibrary{decorations.markings}
\tikzstyle{vecArrow} = [thick, decoration={markings,mark=at position
   1 with {\arrow[semithick]{open triangle 60}}},
   double distance=1.4pt, shorten >= 5.5pt,
   preaction = {decorate},
   postaction = {draw,line width=1.4pt, white,shorten >= 4.5pt}]
\tikzstyle{innerWhite} = [semithick, white,line width=1.4pt, shorten
>= 4.5pt]
\tikzset{
  mid arrow/.style={postaction={decorate,decoration={
        markings,
        mark=at position .5 with {\arrow[#1]{stealth}}
      }}},
}
\tikzset{
    vector/.style={decorate, decoration={snake}, draw},
	provector/.style={decorate, decoration={snake,amplitude=2.5pt}, draw},
	antivector/.style={decorate, decoration={snake,amplitude=-2.5pt}, draw},
    fermion/.style={draw=black, postaction={decorate},
        decoration={markings,mark=at position .55 with {\arrow[draw=black]{>}}}},
    fermionbar/.style={draw=black, postaction={decorate},
        decoration={markings,mark=at position .55 with {\arrow[draw=black]{<}}}},
    fermionnoarrow/.style={draw=black},
    gluon/.style={decorate, draw=black,
        decoration={coil,amplitude=4pt, segment length=5pt}},
    scalar/.style={dashed,draw=black, postaction={decorate},
        decoration={markings,mark=at position .55 with {\arrow[draw=black]{>}}}},
    scalarbar/.style={dashed,draw=black, postaction={decorate},
        decoration={markings,mark=at position .55 with {\arrow[draw=black]{<}}}},
    scalarnoarrow/.style={dashed,draw=black},
    electron/.style={draw=black, postaction={decorate},
        decoration={markings,mark=at position .55 with {\arrow[draw=black]{>}}}},
	bigvector/.style={decorate, decoration={snake,amplitude=4pt}, draw},
}
\def\be{\begin{equation}}
\def\ee{\end{equation}}
\def\bea{\begin{eqnarray}}
\def\eea{\end{eqnarray}}
\def\beal{\begin{align}}
\def\eeal{\end{align}}

\newcommand{\Dzero}{\Delta_{0}}
\newcommand{\Dplus}{\Delta_{+}}

\newcommand{\mchi}{m_{\chi}}

\newcommand{\met}{ $E_T^{\rm miss}$}

\newcommand{\drll}{ $\Delta R(l^+, l^-)~$}
\newcommand{\drllmin}{ $\Delta R_{\rm min}(l^+, l^-)~$}
\newcommand{\metm}{ E_T^{\rm miss}}

\graphicspath{{Plots/}}
\relax

\title{
Probing compressed dark sectors at 100 TeV in
  the dileptonic mono-$Z$ channel }

\author[a]{Rakhi Mahbubani}
\author[b,c]{and Jos\'e Zurita}

\affiliation[a]{Institut de Th\'eorie des Phenom\`enes Physiques, EPFL, Lausanne, Switzerland}
\affiliation[b]{Institute for Nuclear Physics (IKP), Karlsruhe Institute of Technology, Hermann-von-Helmholtz-Platz 1, D-76344 Eggenstein-Leopoldshafen, Germany}
\affiliation[c]{Institute for Theoretical Particle Physics (TTP), Karlsruhe Institute of Technology, Engesserstra{\ss}e 7, D-76128 Karlsruhe, Germany} 
\emailAdd{rakhi@cern.ch}
\emailAdd{jose.zurita@kit.edu}

\abstract{We examine the sensitivity at a future 100 TeV proton-proton collider to compressed dark sectors whose decay products are invisible due to below-threshold energies and/or small couplings to the Standard Model.  
This scenario could be relevant to models of WIMP dark matter,
where the lightest New Physics state is an (isolated) electroweak multiplet whose
lowest component is stable on cosmological timescales.
We rely on the additional emission of a hard on-shell $Z$-boson
decaying to leptons, a channel with low background systematics, and include a
careful estimate of the real and fake backgrounds to this process in
our analysis.
We show that an integrated luminosity of 30 ab$^{-1}$ would allow
exclusion of a TeV-scale compressed dark sector with inclusive production cross
section of 0.3 fb, for 1\% background systematic
uncertainty and splittings below 5 GeV.
This translates to exclusion of a pure higgsino (wino) multiplet with
mass of 500 (970) GeV.
}

\notoc
\begin{document}
\begin{flushright}
CERN-TH/2018-133  \\ TTP18-021 \end{flushright}
\maketitle

\section{Introduction}
\label{sec:intro}
The current state of particle physics brings to mind the (apparently
apocryphal \cite{Wikipedia}) Chinese curse ``May you live in
interesting times''.  That the complete absence of any evidence for
Physics Beyond the Standard Model (BSM) at the LHC is interesting is
undeniable.  What is less clear is how to interpret it: does it signal
a need for a paradigm shift in our understanding of naturalness of
the electroweak scale, or is New Physics simply better concealed
than we had expected?

The latter could occur if the lowest-lying new states are
near-degenerate in mass, leading to decay products that are below
detection thresholds,
particularly if the lightest state in this
compressed spectrum is collider-stable and interacts weakly with
ordinary matter, making it invisible to our detectors.  Production of these compressed states at hadron colliders are thus indistinguishable
from uninteresting Standard Model (SM) background, unless the missing
energy is enhanced by recoil of the invisible system against a hard
visible SM state $X$ which can then be used for triggering and
analysis.

Such `dark
sectors' may be independently motivated by thermal relic dark matter,
as TeV-scale electroweak multiplets with a thermal history have a
density consistent with the measured relic abundance
\cite{Cirelli:2005uq}.\footnote{This would require a large
tuning in the mass of the Higgs, however we find naturalness
considerations misplaced as a motivation for a future
circular collider, touted as a machine of the
`post-naturalness era'~\cite{Golling:2016gvc}.}

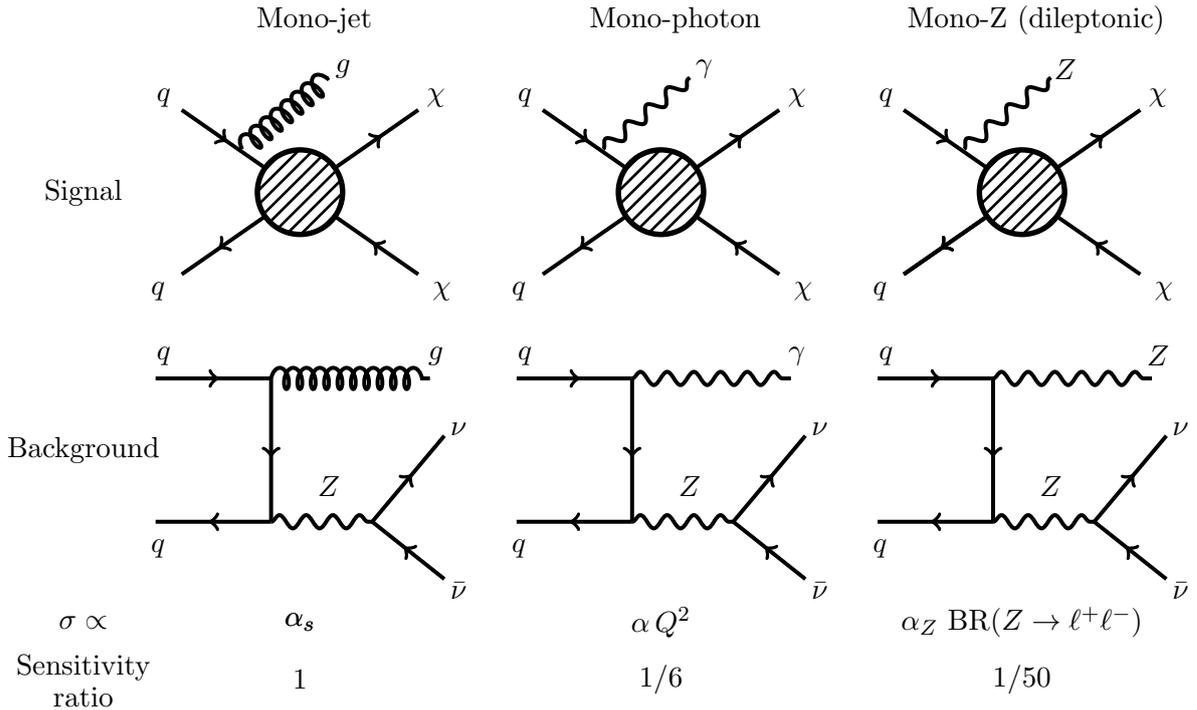
\begin{figure}[!htb]
  \begin{tikzpicture}[line width=1.5, scale=1.9]
    \node at (0.1,1.2) {Mono-jet};
    \node at (-1.5,0) {Signal};
    \draw[fermion](145:1) -- (145:.3cm);
    \node at (145:1.15) {$q$};
    \draw[gluon] (-0.4,0.3)--(0.2,0.8);
    \node at (0.3,0.86) {$g$};
    \draw[fermionbar](215:1) -- (215:.3cm);
    \node at (215:1.2) {$q$};
    \draw[fermionbar](35:1) -- (35:.3cm);
    \node at (35:1.15) {$\chi$};
    \draw[fermion](-35:1) -- (-35:.3cm);
    \node at (-35:1.2) {$\chi$};
    \draw[fill=black] (0,0) circle (.3cm);
    \draw[fill=white] (0,0) circle (.29cm);
    \begin{scope}
      \clip (0,0) circle (.3cm);
      \foreach \x in {-.9,-.8,...,.3}
      \draw[line width=1 pt] (\x,-.3) -- (\x+.6,.3);
    \end{scope}
    \begin{scope}[shift={(2.5,0)}]
      \node at (0.1,1.2) {Mono-photon};
      \draw[fermion](145:1) -- (145:.3cm);
      \node at (145:1.15) {$q$};
      \draw[vector] (-0.4,0.3)--(0.2,0.8);
      \node at (0.3,0.86) {$\gamma$};
      \draw[fermionbar](215:1) -- (215:.3cm);
      \node at (215:1.2) {$q$};
      \draw[fermionbar](35:1) -- (35:.3cm);
      \node at (35:1.15) {$\chi$};
      \draw[fermion](-35:1) -- (-35:.3cm);
      \node at (-35:1.2) {$\chi$};
      \draw[fill=black] (0,0) circle (.3cm);
      \draw[fill=white] (0,0) circle (.29cm);
      \begin{scope}
        \clip (0,0) circle (.3cm);
        \foreach \x in {-.9,-.8,...,.3}
        \draw[line width=1 pt] (\x,-.3) -- (\x+.6,.3);
      \end{scope}
    \end{scope}
    \begin{scope}[shift={(5,0)}]
      \node at (0.1,1.2) {Mono-Z (dileptonic)};
      \draw[fermion](145:1) -- (145:.3cm);
      \node at (145:1.15) {$q$};
      \draw[vector] (-0.4,0.3)--(0.2,0.8);
      \node at (0.3,0.86) {$Z$};
      \draw[fermionbar](215:1) -- (215:.3cm);
      \draw[fermionbar](215:1) -- (215:.3cm);
      \node at (215:1.2) {$q$};
      \draw[fermionbar](35:1) -- (35:.3cm);
      \node at (35:1.15) {$\chi$};
      \draw[fermion](-35:1) -- (-35:.3cm);
      \node at (-35:1.2) {$\chi$};
      \draw[fill=black] (0,0) circle (.3cm);
      \draw[fill=white] (0,0) circle (.29cm);
      \begin{scope}
        \clip (0,0) circle (.3cm);
        \foreach \x in {-.9,-.8,...,.3}
        \draw[line width=1 pt] (\x,-.3) -- (\x+.6,.3);
      \end{scope}
    \end{scope}
    \begin{scope}[shift={(0,-1.8)}]
      \node at (-1.5,0) {Background};
      \draw[fermion](-1,0.5)--(-0.2,0.5);
      \node at (145:1.15) {$q$};
      \draw[fermionbar] (-1,-0.5)--(-0.2,-0.5);
      \draw[fermion] (-0.2,0.5)--(-0.2,-0.5);
      \draw[gluon] (-0.2,0.5)--(0.9,0.5);
      \draw[vector] (-0.2,-0.5)--(0.5,-0.5);
      \draw[fermion] (0.5,-0.5) -- (1,0.1);
      \draw[fermionbar] (0.5,-0.5) -- (1,-0.9);
      \node at (215:1.2) {$q$};
      \node at (35:1.15) {$g$};
      \node at (0.2,-0.25) {$Z$};
      \node at (1.1,0.15) {$\nu$};
      \node at (1.1,-0.98) {$\bar{\nu}$};
      \node at (-1.5,-1.2) {$\sigma\propto$};
      \node at (0,-1.2) {$\alpha_s$};
      \node at (-1.5,-1.52) {Sensitivity};
      \node at (-1.5,-1.72) {ratio};
      \node at (0,-1.2) {$\alpha_s$};
      \node at (0,-1.6) {1};
      \begin{scope}[shift={(2.5,0)}]
        \draw[fermion](-1,0.5)--(-0.2,0.5);
        \node at (145:1.15) {$q$};
        \draw[fermionbar] (-1,-0.5)--(-0.2,-0.5);
        \draw[fermion] (-0.2,0.5)--(-0.2,-0.5);
        \draw[vector] (-0.2,0.5)--(0.9,0.5);
        \draw[vector] (-0.2,-0.5)--(0.5,-0.5);
        \draw[fermion] (0.5,-0.5) -- (1,0.1);
        \draw[fermionbar] (0.5,-0.5) -- (1,-0.9);
        \node at (215:1.2) {$q$};
        \node at (35:1.15) {$\gamma$};
        \node at (0.2,-0.25) {$Z$};
        \node at (1.1,0.15) {$\nu$};
        \node at (1.1,-0.98) {$\bar{\nu}$};
        \node at (0,-1.2) {$\alpha\,Q^2$};
        \node at (0,-1.6) {1/6};
      \end{scope}
      \begin{scope}[shift={(5,0)}]
        \draw[fermion](-1,0.5)--(-0.2,0.5);
        \node at (145:1.15) {$q$};
        \draw[fermionbar] (-1,-0.5)--(-0.2,-0.5);
        \draw[fermion] (-0.2,0.5)--(-0.2,-0.5);
        \draw[vector] (-0.2,0.5)--(0.9,0.5);
        \draw[vector] (-0.2,-0.5)--(0.5,-0.5);
        \draw[fermion] (0.5,-0.5) -- (1,0.1);
        \draw[fermionbar] (0.5,-0.5) -- (1,-0.9);
        \node at (215:1.2) {$q$};
        \node at (35:1.15) {$Z$};
        \node at (0.2,-0.25) {$Z$};
        \node at (1.1,0.15) {$\nu$};
        \node at (1.1,-0.98) {$\bar{\nu}$};
        \node at (0,-1.2) {$\alpha_Z$ BR$(Z\rightarrow \ell^+\ell^-)$};
        \node at (0,-1.6) {1/50};
      \end{scope}
    \end{scope}
  \end{tikzpicture}
  \caption{Relative signal sensitivity in the three main mono-$X$
    channels.  Since the cross sections for each signal and its
    irreducible background scale in the same way, if we assume the
    relative effiency for the signal and background processes are the
    same over each of the different channels, the naive sensitivity, computed as $S/\sqrt{B}$, scales as
    $\sqrt{\alpha_S}:Q\sqrt{\alpha}:\sqrt{\alpha_Z
      \textrm{BR}(Z\rightarrow \ell^+\ell^-)}$.  However this argument
  does not account for systematic uncertainties, which we expect will
  grow at higher centre-of-mass energy.}\label{fig:SensitivityRatios}
\end{figure}

The prevailing wisdom from mono-$X$ searches is that the hadronic
channels (monojet, and hadronic mono-$W$ and mono-$Z$) give the
strongest constraints on pair-production of invisible states, followed
by the monophoton channel, with a far reduced sensitivity acheivable in
leptonic mono-$Z$.  As pointed out in \cite{Bernreuther:2018nat} this trend is
not simply due to the relative production rate for each channel, since
the cross section for the corresponding irreducible
background scales similarly (see Fig.~\ref{fig:SensitivityRatios}).  Hence, for
a search whose background is sufficiently small that its
uncertainty is statistics-dominated, one would naively expect the ratio of significances monojet : monophoton : mono-$Z$ to scale like
$1 :q\sqrt{\alpha/\alpha_s} : \sqrt{{\rm BR}(Z)\alpha_Z/\alpha_s }$ for
$\alpha_i=g_i^2/4\pi$ and ${\rm BR}(Z)$ the branching ratio of the
channel the visible $Z$ boson decays to.\footnote{Here we assumed that the relative efficiencies for the signal and background
processes are similar across the different mono-$X$
channels.  This is not unlikely for a total background dominated by the SM irreducible component, with an on-shell $Z$
  decaying invisibly.}   By this measure the dileptonic mono-$Z$
channel is approximately a fiftieth as
sensitive to compressed dark states as the monojet, and around a tenth as
sensitive as the monophoton.  

Omitted in this argument is the effect of systematics, which is
undoubtedly large due to the large backgrounds.  Moreover, as the
leading (subleading) background in the monojet (mono-$Z$) channel
scale as the gluon Parton Distribution Function (PDF), we would expect
the background systematics to grow with energy. In the regime where the
background uncertainty is dominated by systematic effects there is now
an inherent {\it disadvantage} to using the monojet channel, since the sensitivity
scales like $S/(\beta B)$, for sytematics factor $\beta$,
and the coupling ratios cancel. 

 Of
course this comparison is rather simplistic and a number of additional effects
must be taken into account before
drawing a conclusion: the kinematic suppression due to the production
of the $Z$ boson at LHC energies, as compared with the logarithmic
enhancement of the monojet and monophoton rates, for instance.
This last effect would be partially mitigated at a high-energy hadron
collider like the FCC-hh.
For $\sqrt{s}\gg m_Z$, the $Z$-boson can be thought
of as effectively massless, allowing us to neglect the additional phase
space suppression from production of a massive state, and giving rise
to sudakov divergences which would also need resummation, as for jet and
photon emission.
Moreover the strong coupling runs down at high energies, giving the
monojet channel less of an advantage due to coupling alone, We believe these factors warrant a
reassessment of the
reach for compressed states in the leptonic mono-$Z$ channel, as
compared with that in the monojet channel \cite{Low:2014cba},
at FCC energies.

Existing limits on the reach for compressed dark sectors can be found
in
\cite{Aaboud:2016tnv,Sirunyan:2017hci,ATLAS:2016bza,Sirunyan:2017qfc,Aaboud:2017dor,Sirunyan:2017ewk}.
The reach at hadron colliders has been covered in various studies: e.g
\cite{Gori:2013ala,Han:2013usa,Schwaller:2013baa,Han:2014kaa,Barducci:2015ffa}
consider monojet searches;
\cite{Baer:2014cua,Cirelli:2014dsa,Anandakrishnan:2014exa} look at the
monophoton channel
\cite{Han:1999ne,Alves:2015dya,Bell:2012rg,Anandakrishnan:2014exa}
look at leptonic mono-$Z$.  Other searches rely on the detection of
soft daughter particles from decays within the compressed sector
\cite{Schwaller:2013baa,Han:2014kaa,Low:2014cba}, although their
sensitivity is strongly dependent on $p_T$ reconstruction
thresholds, which will necessarily increase at a future collider.  The
reach due to disappearing charged tracks from a highly-boosted
charged component can be found in
\cite{Cirelli:2014dsa,Low:2014cba,Mahbubani:2017gjh}.  There are in
addition cosmological constraints on pure electroweakino relics,
although the prospects for discovery of a wino-like relic are rather
more promising \cite{Fan:2013faa,Cohen:2013ama}.  Nearly pure
higgsino-like multiplets instead remain elusive
\cite{Kowalska:2018toh}, although recent studies
speculate on the possibility of probing this scenario using observations of compact stars
\cite{Krall:2017xij} or neutron stars \cite{Baryakhtar:2017dbj}.

In this work, we explore the reach at the FCC-hh for compressed dark-sector states
in the dileptonic
mono-$Z$ channel, making careful consideration of both real and fake Standard Model
backgrounds. Our results are given both as contours of production cross section required for exclusion at a given
mass and splitting, for ease of recasting, as well as a reach for a
pure higgsino-like dark sector, which is a simple and compelling example of a
thermal relic.  The cross section limits would hold in many scenarios
with production of a long-lived neutral state, such as some models
of neutrino mass \cite{Gronau:1984ct,Antusch:2015mia,Helo:2018qej}, or
ones containing Higgs portals with tiny couplings
\cite{Curtin:2013fra,Curtin:2015fna}.

Our paper is organized as follows. Section \ref{sec:analysis}
contains specifics of our analysis, with the simplified model used for
simulation purposes detailed in Section \ref{sec:model}.  Event
generation, backgrounds and cuts, including an extended discussion of
shapes of missing transverse energy (MET) distributions is contained in Section
\ref{sec:EvtGen}.  Our results, including some discussion of the
effect of higher-order corrections are shown in Section
\ref{sec:Results}; and we interpret these in the context of a
pure higgsino thermal relic in Section \ref{sec:higgsinoRelic}, which
will also be constrained by future indirect- and direct-detection
experiments.  We conclude in Section \ref{sec:conclu}.  
\section{Mono-Z analysis}
\label{sec:analysis}
\subsection{Simplified model}
\label{sec:model}

For the purposes of simulation and analysis, we focus on the case
with a dark sector consisting of a weak doublet
$\chi=(\chi^+,\chi^0)$, with
hypercharge $Y= -1/2$ and Dirac mass $m_\chi$.
We include tree-level mixing effects
with a heavy electroweak-singlet fermion with Majorana mass $m_S$, as shown in Eq.(\ref{equ:Lagrangian})
\be
{\cal L } \supset i \bar{\chi}  \not \negthickspace D  \chi
+\frac{i}{2} \bar{\lambda} \not \negthickspace \partial \lambda -
m_\chi \bar{\chi} \chi - \frac{m_S}{2} \bar{\lambda} \lambda - y_L
\bar{\lambda}H^\dag P_L \chi + y_R \bar{\lambda}H P_R \chi +
\textrm{h.c.} \,
\label{equ:Lagrangian}
\ee
We take all parameters to be real\footnote{Although the system contains one
irreducible phase \cite{Mahbubani:2005pt} this has no effect on the collider
  phenomenology.} and fix the yukawa couplings to the corresponding
  values in the neutralino sector of the Minimal Supersymmetric
  Standard Model (MSSM) for concreteness, with
  $\tan(\beta)=15$.  We focus on the experimentally-challenging regime
  of large $m_S$, which corresponds to the MSSM in the pure higgsino
  limit.  In this regime the heaviest neutral state $\chi_3^0$, which is almost pure
singlet, is irrelevant for the collider phenomenology, and can be
ignored.  Moreover, after electroweak symmetry breaking the mass
mixing with the singlet splits the neutral Dirac fermion into two
Majorana fermions $\chi^0_1$ and $\chi^0_2$; for large $m_S$ the mass
splitting between these two states, $\Delta_0=\chi^0_2-\chi^0_1$ is too small to give
rise to detectable decay products:
\begin{equation}
\Delta_0\sim 200\,\textrm{MeV} \left(\frac{10\,\textrm{TeV}}{m_S}\right)\,.
\end{equation}
This is also the case for the charged-neutral splitting
$\Delta_+=m_{\chi^+}-m_{\chi^0}$, which gets an additional
contribution from electroweak loops of $\Delta_\textrm{1-loop}\sim\mathcal{O}(\alpha\,m_Z)\sim
300$ MeV \cite{Thomas:1998wy}. 

Rather than the Lagrangian input parameters, we will will express our results in terms of the
phenomenologically-relevant parameter set ($m_\chi,\Delta_+$), using the
latter as a proxy for the average mass splitting within the dark
sector.  Our search sensitivity will be
strongly dependent on $m_\chi$, which fixes the production cross
section for a given electroweak representation (in the small-mixing limit),
and also determines the kinematics of the events.  By contrast the sensitivity has
a weak dependence on $\Delta_+$, through the cut efficiencies,
and only for
$\Delta_+$ above a certain threshold, beyond which the visible decay products have enough energy to fall foul
of our object vetos.\footnote{In this regime the contribution to the splitting due to electroweak loop
  effects are sub-dominant.}  Below this threshold, which we see is
around 5 GeV, ours is a
one-parameter analysis which is completely insensitive to the details
of the decays.  As such we can apply the limits on the total production
cross section obtained in this scenario as a conservative estimate on
the sensitivity in this channel to arbitrary compressed dark sectors,
produced in association with a leptonically-decaying $Z$, provided the
dark sector can be approximately characterized by a single mass
scale.

If the lightest neutral component of the multiplet, $\chi_1^0$, is stable on
cosmological timescales, it becomes a good candidate for dark matter, saturating the  relic density at a mass of $1.1$ TeV \cite{Giudice:2004tc,Cirelli:2005uq}.
For a relic $\chi$, the presence of a tree-level splitting $\Delta_0$ due to mixing with the
heavy singlet is essential for it not to be ruled out by direct
detection experiments, a vector-like coupling between dark matter and
the SM $Z$-boson being long-excluded due to too large
a scattering cross section with ordinary matter.

\begin{figure}
\centering
\includegraphics[width=0.7\textwidth]{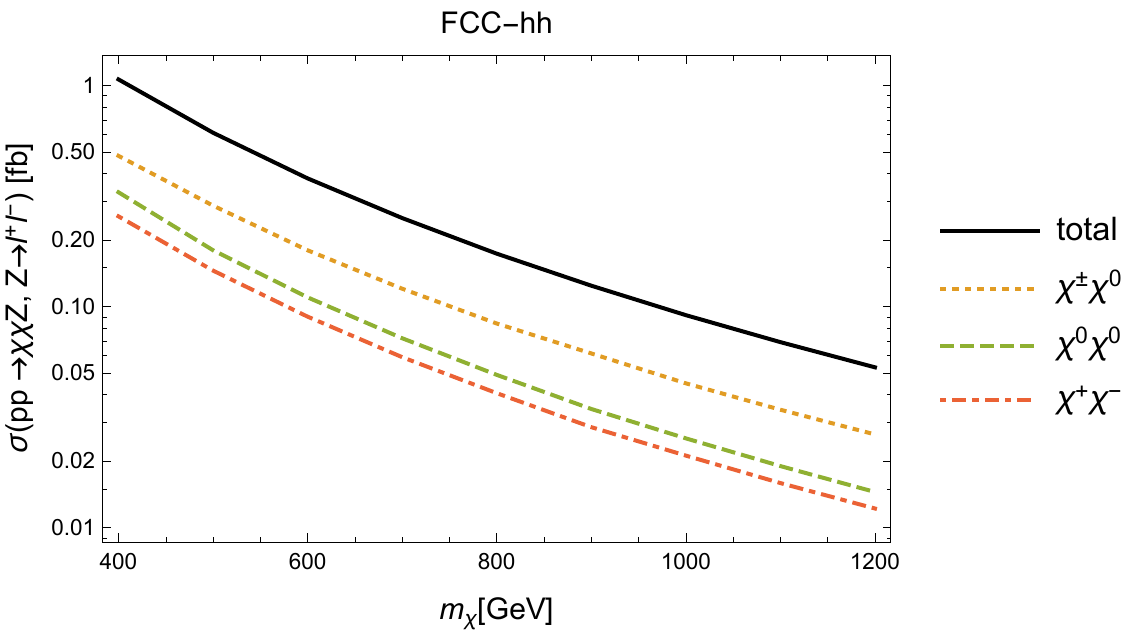}
\caption{Inclusive production cross sections for $p p \to \chi \chi Z\,, (Z\to\ell^+\ell^-)$, for
  weak-doublet Dirac fermion $\chi$ with hypercharge -$1/2$ at a
  future hadron-hadron collider with 100 TeV centre-of-mass energy.
  We assume that inter-state splittings are small enough to have a
  negligible effect.}
\label{fig:incXS}
\end{figure}
We show in Fig.~\ref{fig:incXS} the inclusive production cross section for
pair-production of a weak-doublet $\chi$ with hypercharge -$1/2$, in association with a leptonically-decaying
$Z-$ boson, for negligible inter-state splittings.  This ranges from 1
fb for a mass of 400 GeV to 0.05 fb at 1.2
TeV. 

\subsection{Event generation}
\label{sec:EvtGen}

We manually implemented the simplified model above via the \textit{usr\_mod} feature
in \texttt{MadGraph5 v2.3.3}~\cite{Alwall:2014hca}, and
tested it against existing MSSM implementations. Widths and branching fractions were computed using analytic
expressions given in Appendix A of \cite{Mahbubani:2017gjh} (and
references therein), and input manually, since existing SUSY spectrum
generators do not cover the regime where tree-level mixing is small and the
masses of the SM fermion decay products are relevant.

Our signal consists of the production of a pair of electroweakinos
recoiling against a hard leptonically-decaying $Z$-boson, giving rise to
two leptons plus missing transverse energy in the final state.  We
simulate leading-order $\chi\bar{\chi}Z$ production, and the leptonic decay of the $Z$
boson, in \texttt{MadGraph5
  v2.3.3} with the CTEQ6L1 parton
distributions~\cite{Pumplin:2002vw}, and allow \texttt{Pythia
  v6.4}~\cite{Sjostrand:2006za} to handle dark sector decays, parton showering and hadronization.
We use \texttt{Delphes v.3.2.0}~\cite{deFavereau:2013fsa} for detector
simulation, with an FCC-hh card that is customized to 
impose no detector-level lepton isolation in order to maximize our
sensitivity to highly-boosted $Z$-bosons.  We instead impose a
  naive analysis-level isolation-based rejection of leptons that are
  within $\Delta R<0.2$ of a jet. (See object selection below.)  Jets
are clustered with FASTJET's \cite{Cacciari:2011ma} anti-kt algorithm
~\cite{Cacciari:2008gp} for a jet radius of R=0.4.
Events were read using
\texttt{MadAnalysis5}~\cite{Conte:2012fm,Conte:2014zja} and analysed
using in-house code.  

Backgrounds, which were also simulated using the above pipeline, can be split into three categories:

\begin{itemize}
\item{real backgrounds}: Processes giving rise to two leptons and
  missing energy at parton level. This includes the diboson processes
  $Z Z \to l^+ l^- \nu \nu$ and $W^+ W^- \to l^+ \nu l^- \nu$, as well as
  fully-leptonic $t \bar{t}$. These are simulated at leading order.

\item{one fake/lost lepton}: Processes that have less than (more than)
  two hard leptons at parton level, requiring the mis-identification
  of jets as leptons (one or more leptons to be missed). These
  include semi-leptonic $t \bar{t}$ (merged and matched up to one
  additional jet), (leptonic) $W$ + jets (merged and matched up to
  two jets) and fully-leptonic $WZ$  (simulated at parton level).

\item{fake MET}: The missing energy in these processes arises mainly
  from mis-measurement of the transverse momenta of hard jets. This
  category includes $Z \to l^+ l^-$ + jets (matched up to two jets), and also
  the diboson processes $ZZ$ and $ZW$, with one $Z$ decaying
  leptonically and the other gauge boson decaying hadronically
  (simulated at parton level). 

\item{fake leptons and fake MET}: The leading contribution to this category
  of background events would come from QCD multijets, which has a huge
  cross section.  In
    principle we would expect an increased contribution from multijets
    to the total background at the FCC-hh as compared with the LHC, since this process grows with the square
    of the gluon PDF.  It is however difficult to estimate its
    contribution due to generation efficiencies, and we assume this
    will be negligible after our selection cuts, as it is in existing
    mono-$Z$ analyses at the LHC.

\end{itemize}

Generator-level cuts are imposed on the missing energy (and proxies
thereof) in order to improve the efficiency of
both signal and background generation.  For any process with a real
leptonically-decaying $Z$-boson and $\metm$, we impose a hard cut on the transverse momentum
of the $Z$, $p_T(Z)> 400$ GeV.  For all other backgrounds with real $\metm$, we
impose $\metm > 400$ GeV at parton level, while backgrounds with fake
$\metm$ have a hard cut on  $H_T = \sum_{\rm jets}  | p_T | > 400 $ GeV as a
proxy for the maximum $\metm$ in the process.  The value chosen for
this cut is sufficiently below our initial selection so as not to affect
the final cross section.

Our object selection is as follows:

\begin{itemize}
\item $p_T > $  100 (60) GeV for jets (leptons);
\item for leptons: $\Delta_R (l,j) < 0.2$.  We exclude any leptons that
  do not satisfy this criterion.
\end{itemize}

We pre-select events with:
\begin{itemize}
\item exactly two opposite sign, same flavour leptons, which
  reconstruct an on-shell $Z$, $M_{l^+ l^-} \in [76,106]$ GeV;
\item $p_T(Z) > 450$ GeV;
\item $\Delta{\phi}(j_{1,2},\metm)\ge 0.2$ to reduce jets faking $\metm$;
\end{itemize}

Our events consist of three
independent elements that recoil against each other in the transverse
plane: a reconstructed $Z$, \met, and one or more
additional jets, with total transverse momentum $\sum_{\rm
    jets} \vec p_T$.
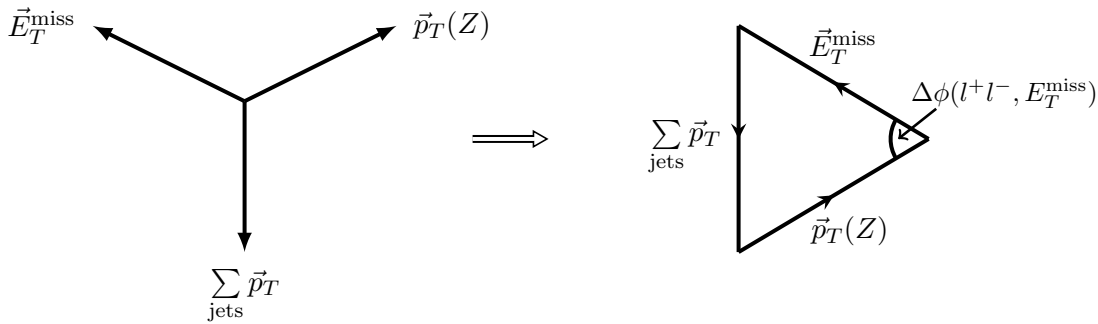
\begin{figure}[!htb]
\begin{tikzpicture}
\draw [->, line width = 1.5pt, >=latex] (0,0) -- (2,1) node [inner sep = 2mm, right] {$\vec p_T(Z)$};
\draw [->, line width = 1.5pt,   >=latex] (0,0) -- (-2, 1) node [inner sep = 2mm, left] {$\vec E_T^{\rm miss}$};
\draw [->, line width = 1.5pt, >=latex] (0,0) -- (0,-2) node [inner
sep = 2mm, below] {$\sum\limits_{\rm jets} \vec p_T$};
\draw[vecArrow] (3,-0.5) -- (4,-0.5);
\path[draw, line width = 1.5 pt, postaction={mid arrow = black}]
(6.5,-2.0) -- (9.0,-0.5) node  [inner sep = 2mm,
below left = 0.8 cm and 0.3cm] {$\vec p_T(Z)$};
\path[draw, line width = 1.5 pt, postaction={mid arrow = black}]
(9,-0.5) -- (6.5,1.0) node [inner sep = 2mm,
below right = -0.2 cm and 0.7 cm] {$\vec E_T^{\rm miss}$};
\path[draw, line width = 1.5 pt, postaction={mid arrow = black}]
(6.5,1.0) -- (6.5,-2.0) node [inner sep = 2mm,
above left = 0.8 cm and 0 cm] {$\sum\limits_{\rm
    jets} \vec p_T$};
\node at (9.0,-0.5) (a) {};
\node at (10.0,0.1) (ap) {\small $\Delta
  \phi (l^+ l^-, \metm)$};
\node at (6.5, 1.0) (b) {};
\node at (6.5, -2.0) (c) {};
\draw [->, bend left, line width = 1.0 pt] (9.1,-0.1) -- (8.6,-0.5);
\pic [draw, line width = 1.5 pt, angle eccentricity = 1.5] {angle = b--a--c};
\end{tikzpicture}
\caption{Sketch of different components of $Z+\!\!$\met$\,$ events
  in the transverse plane: the dilepton system recoils against the
  missing energy and the vector sum of hadronic $p_T$.  Conservation of transverse
  momentum implies the vectors form a closed triangle, which can be
  defined by the lengths of two sides and the enclosed
  angle, or by the lengths of all three sides.}
\label{fig:eventDia}\end{figure}
The relative proportions of these three components vary between the
signal and backgrounds processes, with conservation of transverse momentum requiring
that the three vectors form a closed triangle, see Fig.~\ref{fig:eventDia}.  Hence the transverse
kinematics of any event are completely fixed by specifying the lengths
of the three sides of the triangle, or equivalently the lengths of two
sides and the enclosed angle.  Signal events will mostly resemble an
isosceles triangle, with a hard $Z$ recoiling against missing energy, and
some additional soft hadronic activity.   Backgrounds with fake MET,
on the other hand, will tend more towards equilateral in shape. 

For universality 
across different dark sector mass scales, we re-express these three
dimensionful quantities as two dimensionless ratios, $p_T(Z) / \metm$
and $H_T/\metm$, for $H_T=\sum_{\rm jets}\left|\vec p_T\right|$, with the magnitude of the missing energy setting the
overall mass scale.  We gain in sensitivity by using the scalar sum, $H_T$, rather
than the vector sum of jet transverse momenta (or equivalently $\Delta
  \phi (Z, \metm)$), since the former is also sensitive to
  back-to-back jets, and hence is a better measure of total hadronic
  activity.  Moreover, $H_T$ and $\metm$ share some
  systematics related to jet-mismeasurement, which would cancel to a
  large extent in the ratio. Note that we use 
    $H_T$ as computed by Delphes using
    reconstruction-level jets, which is unaffected by our hard jet
    selection above.  

\begin{figure}[!htb]
	\centering
 	\includegraphics[width=0.49\linewidth]{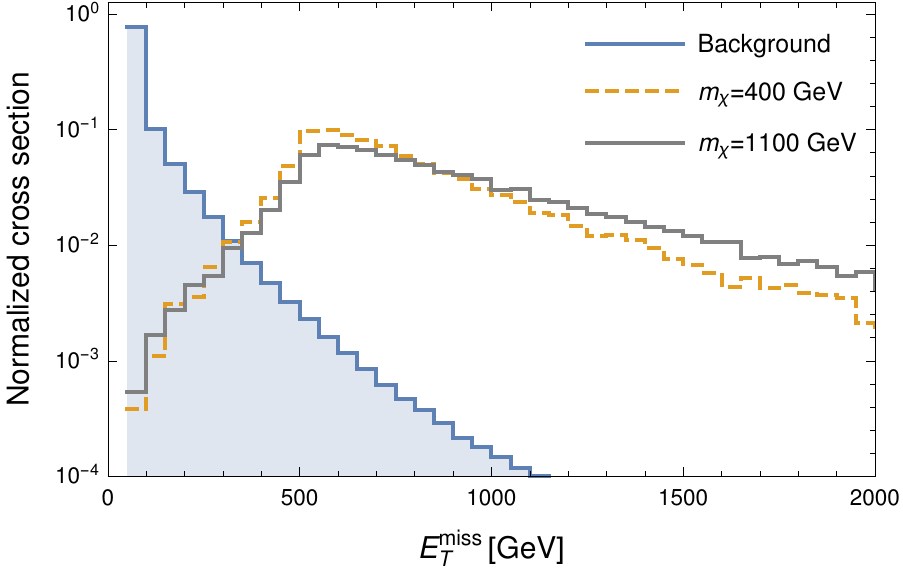}
	\includegraphics[width=0.49\linewidth]{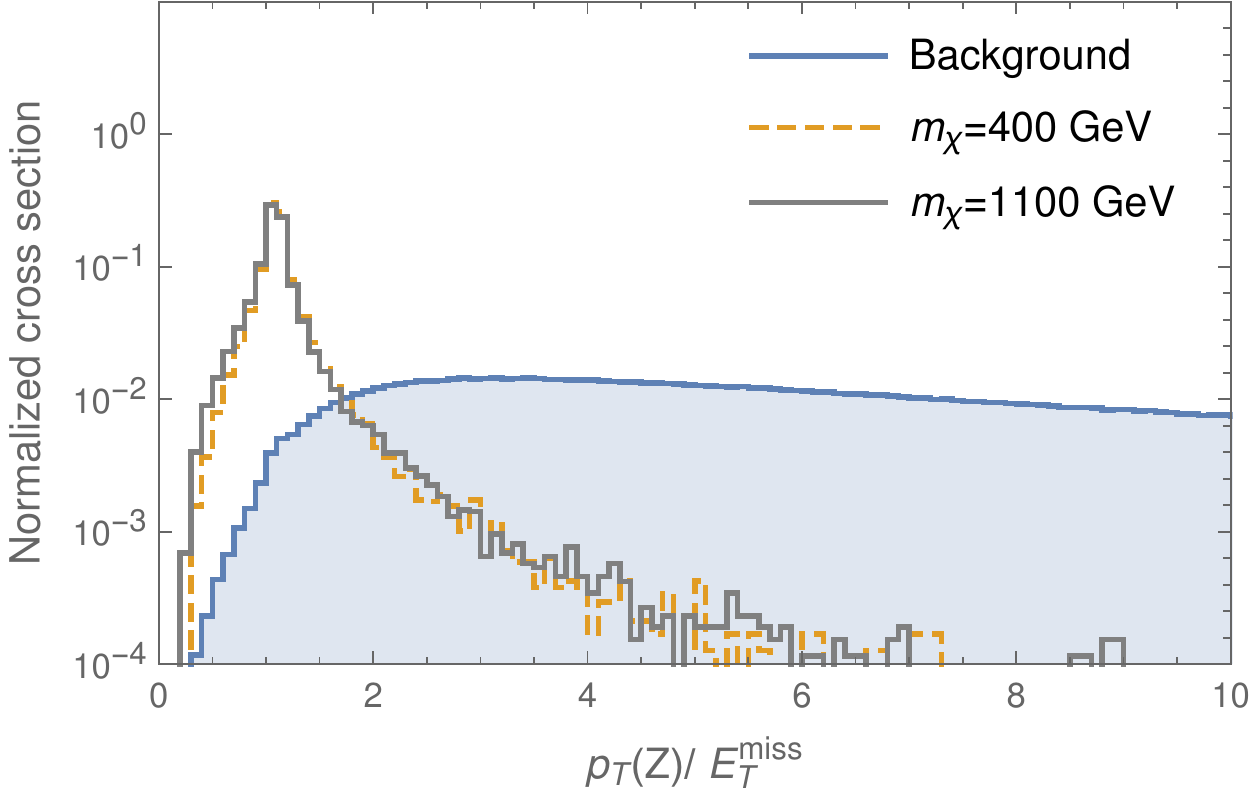}
	\includegraphics[width=0.49\linewidth]{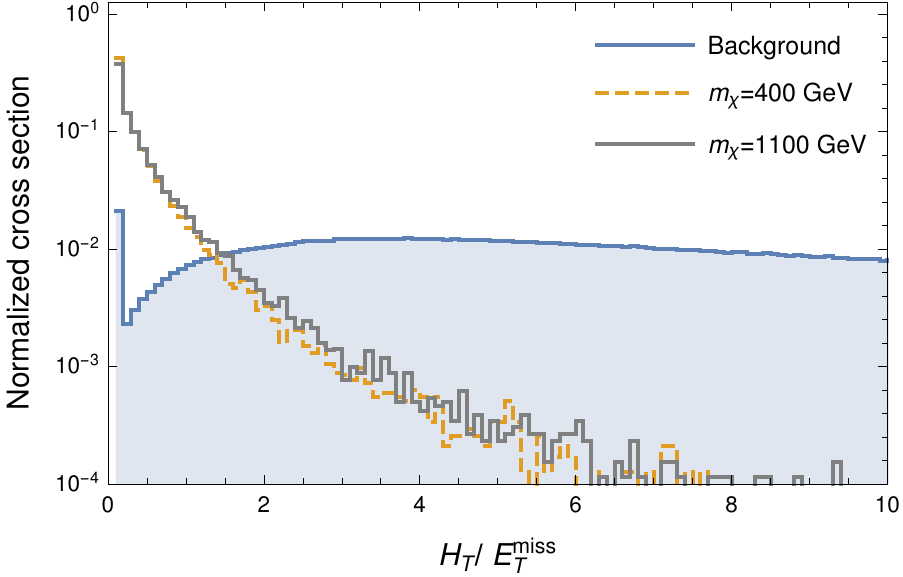}
 	\includegraphics[width=0.49\linewidth]{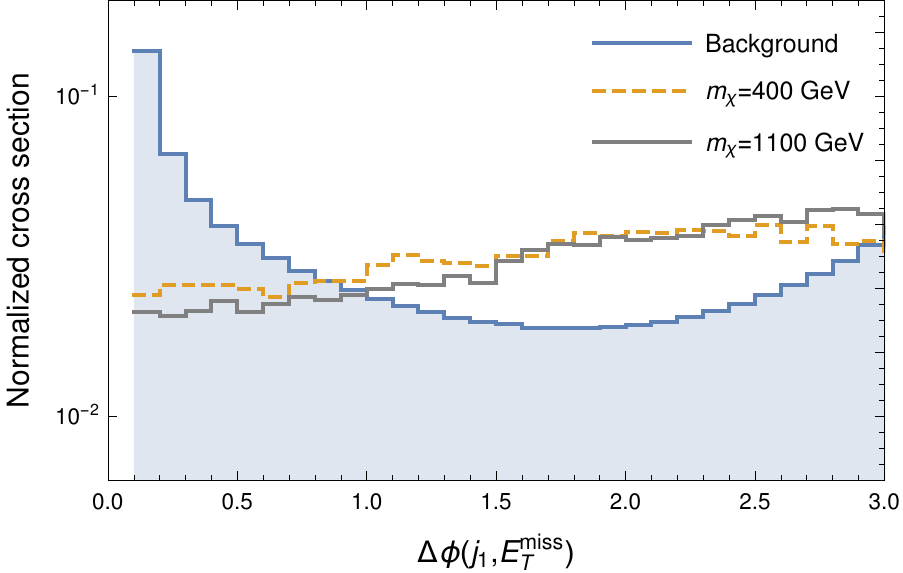}
	\caption{Unit-normalized kinematic distributions after preselection for signal points
          with masses 1.1 TeV (solid) and 400 GeV (dashed),
          and nominal charged-neutral splitting, and the sum of the three leading backgrounds (shaded)
          ($Z+$jets, $ZZ \to l^+ l^- j j$ and $WZ  \to
          \ell^+\ell^-\ell\nu$).}
	\label{fig:Bunitdistros}
	\end{figure}	

Cuts on variables outside the transverse plane, such as lepton
pseudorapidities \cite{Alves:2015dya}, or angular variables in the rest frame of the
$Z$ \cite{Yang:2017iqh} are ineffective here, since the
production mechanisms of the signal and irreducible background are
identical. 

In Fig.~\ref{fig:Bunitdistros} below we display normalized
distributions of relevant kinematic variables after preselection cuts, for dominant backgrounds
and two signal parameter points: $m_\chi=400,\,1100$ GeV with
negligible splitting. The latter point saturates the thermal relic
density measurement (see Section \ref{sec:higgsinoRelic} for
details).  We see that there is significant separation between signal
and background shape in the variables, with the different signal
distributions looking
rather similar over the entire mass range of interest.

Based on these distributions we add to the preselection a preliminary soft $\metm$ cut 

\begin{itemize}
\item $\metm \ge 450$ GeV,
\end{itemize}

and optimize the signal sensitivity (defined as $S/\sqrt{B}$) with
respect to the dimensionless kinematic variables shown.  We find
that the sensitivity is maximized
by imposing the following additional cuts on signal and backgrounds:
\begin{itemize}
\item $x\equiv p_T(Z) / \metm < 1.3$;
\item a hard cut on hadronic activity, $y\equiv H_T/\metm < 1.4$.
\end{itemize}
Optimizing individually for each parameter point yields efficiency differences
that are below 0.1\%.  We further impose a floating $\metm$ cut that is optimized for each $m_\chi$.

\section{Results}\label{sec:Results}

We display the cutflow for signal points with $m_\chi=1.1$ TeV (Signal
A) and $m_\chi=400$ GeV (Signal B) with nominal
splittings, and all relevant backgrounds in Table
\ref{tab:cutflowRM}. The optimal missing energy cut for benchmark $A$
($B$), in the absence of systematic uncertanties, is
$\metm > 900~(550)$ GeV.  We include signal-to-background ratios,
$S/B$ and significances, both with and without systematic
uncertainties, without re-optimizing.  The significance including
systematics is computed as $S/\sqrt(B+\beta^2 B^2)$ for $\beta=$1\%.
\begin{table}[tb!]
	\centering
	\scriptsize
	\input{tables/cutflow}
	\caption{Cut flow for the backgrounds and for signal points
          with with masses 1.1 TeV (A) and 400 GeV (B),
          and nominal charged-neutral splitting (optimized for zero
          background systematics in each case). The numbers of events quoted correspond to a
          total integrated  luminosity of $30$~ab$^{-1}$ at a 100 TeV
          center-of-mass energy. The significance is
          computed assuming a) no systematic errors and b) 1 $\%$
          systematic errors.}
	\label{tab:cutflowRM}
\end{table}

We see from Table \ref{tab:cutflowRM} that while $Z+$jets is the dominant source of background
before optimization, after all cuts its proportion
decreases to around 30\% of the total, with the irreducible di-$Z$ process
accounting for most of the remainder. Sub-dominant backgrounds include
$ZW\to \ell^+\ell^-\ell\nu$ and top quark production in the
semileptonic and pure leptonic channels.  The background composition 
is consistent with that in the 13 TeV, 2.3 fb$^{-1}$ CMS study
\cite{Sirunyan:2017onm}.\footnote{The most recent analyses by CMS 
\cite{Sirunyan:2017qfc} and ATLAS \cite{Aaboud:2017bja} have the
proportion of $Z+$jets decreasing to
around 5\% of the
total background due to a hard cut in $\Delta
  \phi (Z, \metm)$.  However this is a result we are unable to achieve in our
100 TeV analysis without cutting away our signal entirely, as the
proportion of $V+$jets in the background increases with the gluon PDF.}  Neglecting the $Z+$jets background would lead
to an increase of 15-20\% in significance for our signal points.
Note that these sensitivities are contingent upon our ability to
successfully reconstruct the $Z$ using highly-boosted leptons.  The
detrimental effect on the significance of a lower limit on the lepton
separation that can be resolved is shown in appendix \ref{sec:leptonres}.

Clearly the main issue in this analysis is the huge irreducible di-$Z$
background, whose distributions closely mirror those of the signal.  Conventional wisdom holds that a hard enough
cut in missing energy can help suppress this background to reasonable
levels, the signal having a larger fraction of events at high MET.
This is not the case here, where the larger available centre-of-mass energy means
we can access energy scales that are much larger than any relevant mass scale. Any difference in the shape of the MET spectra
of the signal and irreducible background is limited to the region
below the dark sector mass scale; far above this scale both the
$Z$-boson and dark sector particles are effectively massless, hence cutting at
arbitrarily high $\metm$ does not enhance the signal
sensitivity.   We can gain some further understanding of the power of
the MET spectrum by considering the phase space and production
topologies for the signal and irreducible background in MET+$X$
processes. 
 
\subsection*{Interlude: MET shapes}\label{sec:METshape}

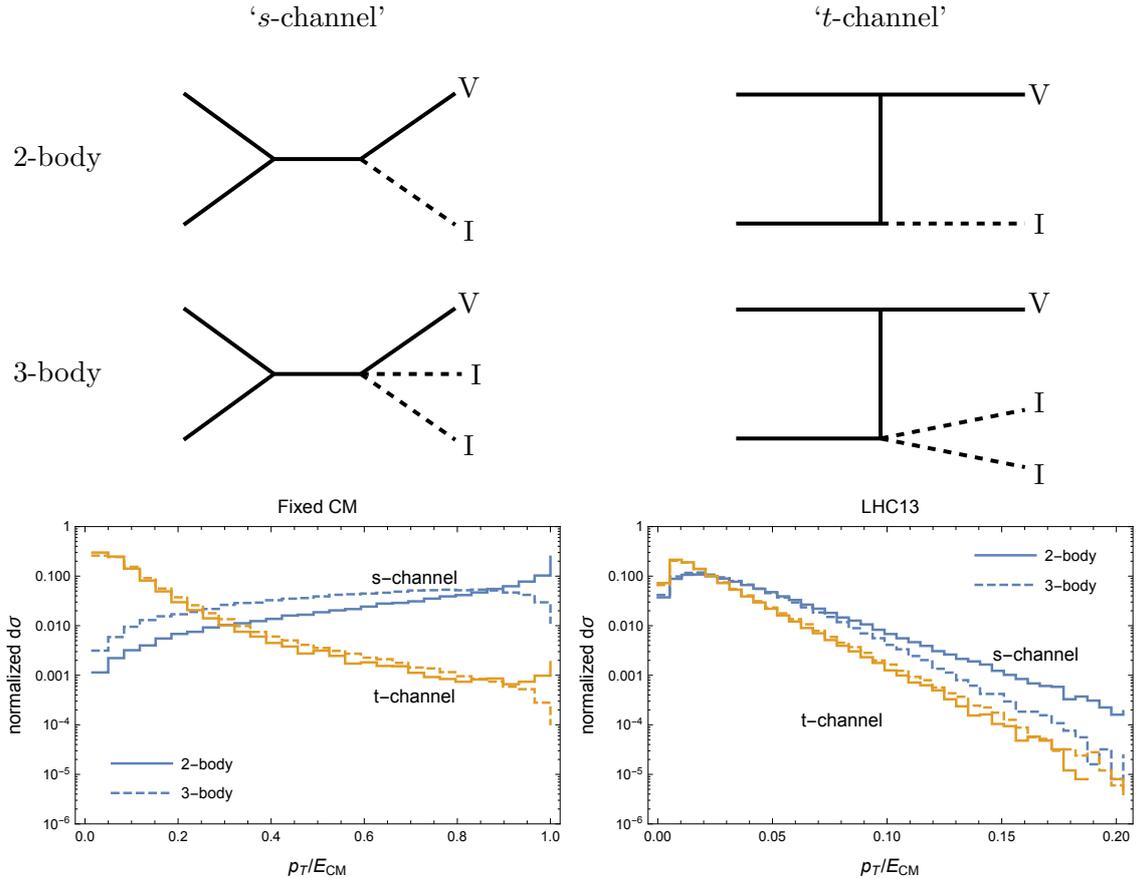
\begin{figure}[tb!]
  \begin{tikzpicture}[line width=1.5, scale=1.9]
    \node at (0.3,1) {`$s$-channel'};
    \node at (-1.5,0) {2-body};
    \draw[-] (0,0) -- (-0.6255,0.459);
    \draw[-] (0,0) -- (-0.6255,-0.459);
    \draw[-] (0,0) -- (0.6,0);
    \draw[-] (0.6,0) -- (1.255,0.459);
    \node at (1.35,0.5) {V};
    \draw[dashed] (0.6,0) -- (1.255,-0.459);
    \node at (1.35,-0.5) {I};
    \begin{scope}[shift={(4.2,0)}]
      \node at (0,1) {`$t$-channel'};
      \draw[-](-1,0.45)--(0,0.45);
      \draw[-] (-1,-0.45)--(0,-0.45);
      \draw[-] (0,0.45)--(0,-0.45);
      \draw[-] (0,0.45)--(1,0.45);
      \draw[dashed] (0,-0.45)--(1,-0.45);
      \node at (1.1,-0.45) {I};
      \node at (1.1,0.45) {V};
    \end{scope}
    \begin{scope}[shift={(0,-1.5)}]
      \node at (-1.5,0) {3-body};
      \draw[-] (0,0) -- (-0.6255,0.459);
      \draw[-] (0,0) -- (-0.6255,-0.459);
      \draw[-] (0,0) -- (0.6,0);
      \draw[-] (0.6,0) -- (1.255,0.459);
      \node at (1.35,0.5) {V};
      \draw[dashed] (0.6,0) -- (1.255,-0.459);
      \node at (1.4,0) {I};
      \draw[dashed] (0.6,0) -- (1.3,0);
      \node at (1.35,-0.5) {I};
      \begin{scope}[shift={(4.2,0)}]
        \draw[-](-1,0.45)--(0,0.45);
        \draw[-] (-1,-0.45)--(0,-0.45);
        \draw[-] (0,0.45)--(0,-0.45);
        \draw[-] (0,0.45)--(1,0.45);
        \draw[dashed] (0,-0.45)--(1,-0.25);
        \draw[dashed] (0,-0.45)--(1,-0.65);
        \node at (1.1,-0.2) {I};
        \node at (1.1,-0.7) {I};
        \node at (1.1,0.5) {V};
      \end{scope}
    \end{scope}
  \end{tikzpicture}
  \includegraphics[width=0.49\linewidth]{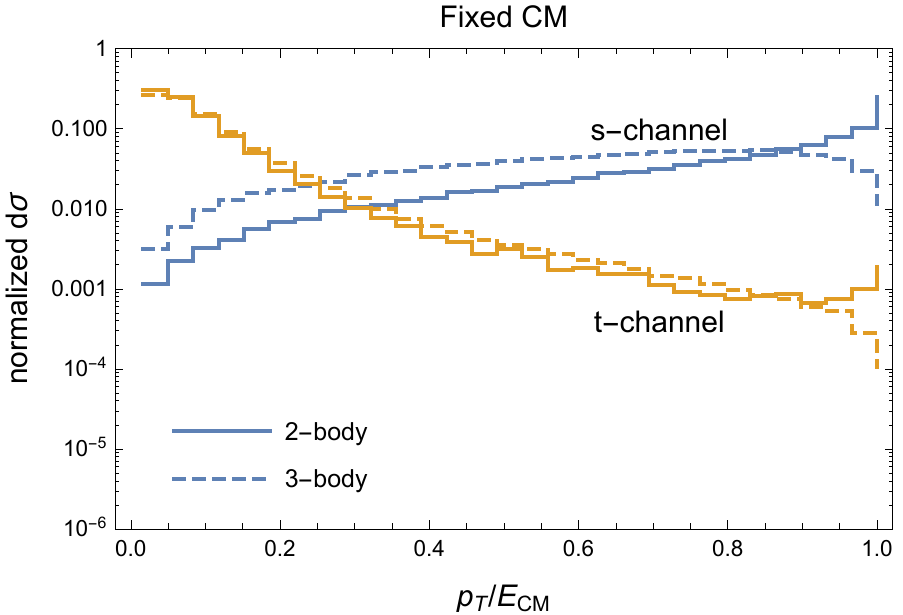}
  \includegraphics[width=0.49\linewidth]{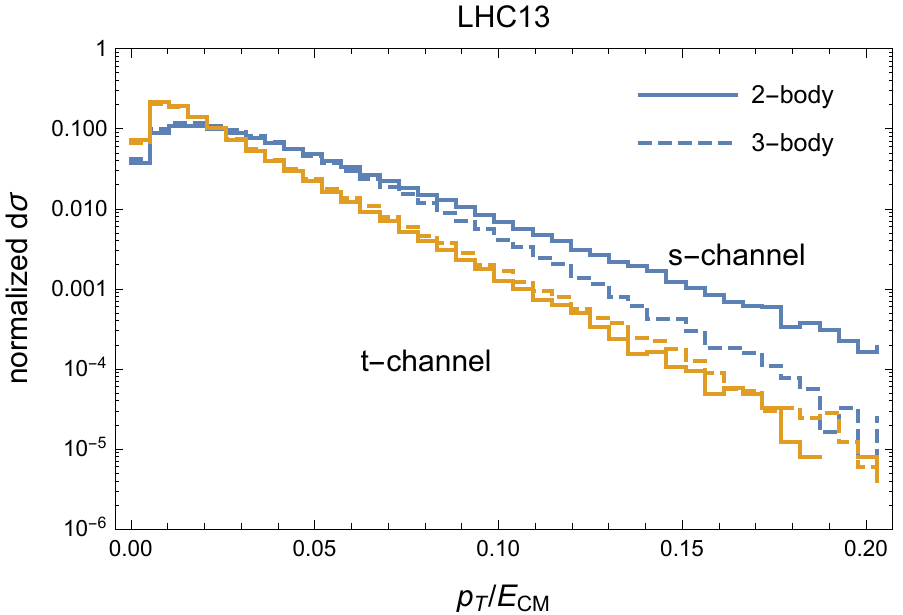}
  \caption{Normalized MET distributions for `$s$-channel' and
    `$t$-channel', 2- and 3-body topologies, with visible final state
    $V$ and invisible states $I$.  The 2-body $t$-channel topology
    corresponds to the irreducible SM di-$Z$ background, as well as
    $\chi\bar{\chi}Z$ production via an on-shell mediator.
    A compressed dark state produced instead via an off-shell
    gauge boson mediator, as it is here, is a 3-body, $t$-channel process.}\label{fig:METShape}
\end{figure}
First, consider the MET spectrum from pure phase space
considerations (which is also the spectrum for production processes
with an $s$-channel propagator).  At fixed center-of-mass energy
the MET distribution for a two-body semi-visible final state has a
jacobian peak at its endpoint, familiar from the transverse mass
distribution of the $W$ boson.  The visible $p_T$ distribution is
instead significantly
softened for a three-body process with one visible final-state particle, since we must integrate over a
larger number of un-measured (invisible) momenta
\cite{Giudice:2011ib}.  A similar pattern must also be evident, although less obvious, at a hadron collider, where
the pure phase space distributions are convoluted with a falling
PDF. Although there is now no jacobian peak, the MET spectrum for the
two-body process is still harder than that for the three-body one.  
This effect is illustrated in
Fig.~\ref{fig:METShape}, where the distributions were
generated using Madgraph simulations of a simplified model with
multiple scalars. If
our signal and irreducible background were produced via this
$s$-channel topology, the latter would have a harder MET spectrum,
since it is a two-body process, and cutting on MET would be counterproductive.

Instead both the three-body signal process $\chi\bar{\chi}Z$ and the
two-body background $ZZ$ are produced in the `$t$-channel', defined
in the right-hand panel of Fig.~\ref{fig:METShape}.\footnote{Note that in the three-body topologies an
offshell particle connecting the $s$- and $t$-channel propagators to
the two invisible states has been omitted for simplicity, since its
presence has no impact on the MET shape.}
Production via a $t$-channel propagator favours forward
emission, as pointed out in \cite{Neubert:2015fka}.  This can mask the phase
space effect, resulting in two-
and three-body $p_T$ spectra that are much more similar in shape (and
softer than the $s$-channel case) both at
fixed and variable centre-of-mass energies.  This argument also holds
for the mono-$W$, monophoton and mono-higgs channel. In the monojet
case it is further complicated by the
fact that the final state gluon can also be crossed into the initial
state, giving a final MET spectrum that is some combination of $t$-
and $s$-channel production.  

This also means that any MET+$X$ search will be maximally sensitive to
a dark sector where $\chi\bar{\chi}X$ can be produced dominantly from an
$s$-channel process via an on-shell dark mediator.  It would be a fun
exercise to find a viable model in which this occurs, and recast current MET+$X$ limits in the context of such a
topology.

\subsection*{Binned MET exclusion}

Even though the irreducible background could in principle be estimated
precisely using measurements of related di-boson processes, as in \cite{Lindert:2017olm},
even small systematic effects can completely swamp our
signal, especially at low masses where backgrounds are large, leaving
no 
exclusion in the mass range of interest.  It is particularly
beneficial in such a situation to move to a binned analysis,
where we can benefit maximally from shape differences between the MET
distribution for the signal and total background (see Fig.~\ref{fig:Bunitdistros}, first
panel).  This minimizes the effect of background systematics, which in
the case where the uncertainties are uncorrelated in different bins, 
would dominantly affect the background normalization rather than its
shape, and is the method
of choice in the most recent $\metm+X$ searches at the LHC (see
e.g. \cite{Sirunyan:2017hci}). 

We use the profile likelihood method to
compute the 95\% CLs exclusion on the total inclusive
$\chi\chi+Z,\,(Z\to\ell^+\ell^-)$ cross section, $\sigma^{95}$.  We
bin both signal and background using 200-GeV $\metm$ bins, after
preselection and $x$ and $y$ cuts, which largely
eliminate the
fake background, and our results are independent of the binning for
reasonable bin widths.   Details of our method are deferred to
Appendix \ref{app:ProfileLikelihood}.  

\begin{figure}[!htb]
	\centering
        \includegraphics[width=0.49\linewidth]{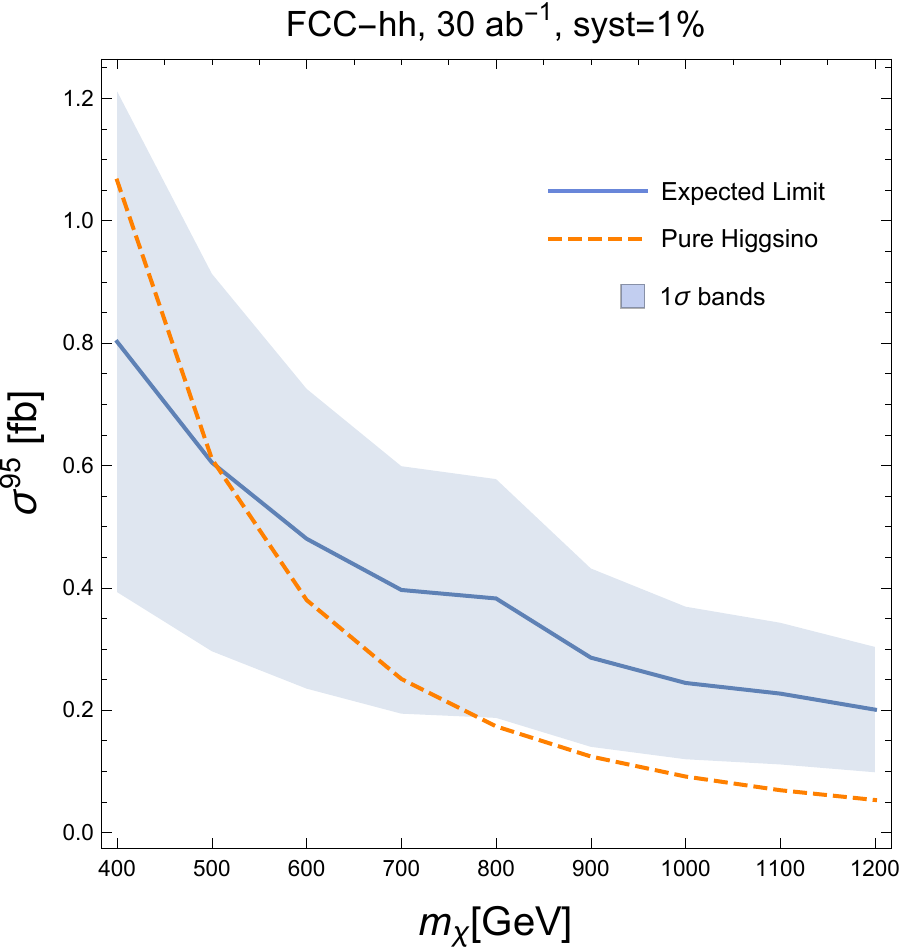}
	\includegraphics[width=0.49\linewidth]{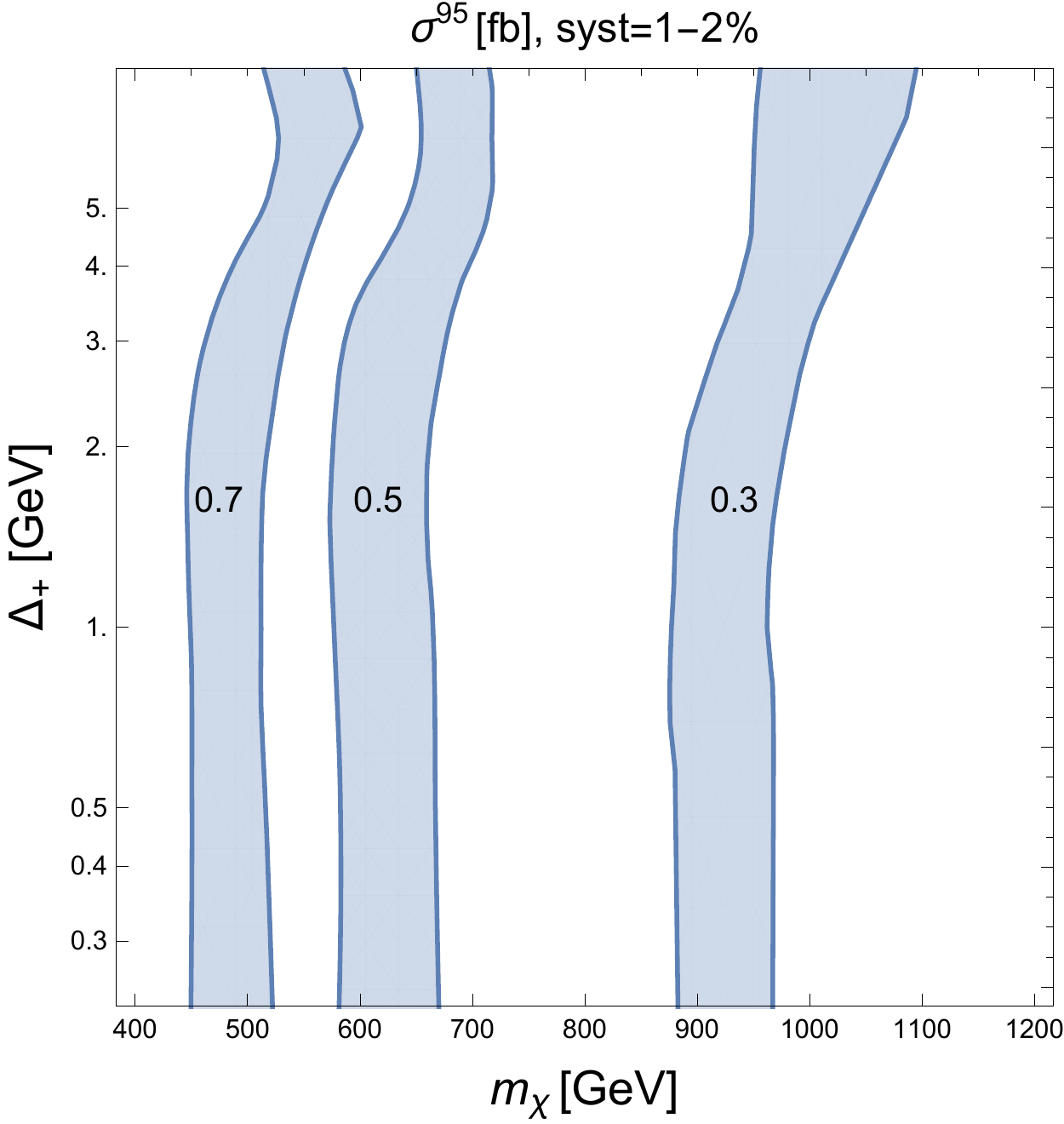}
	\caption{(Left panel) Median expected limit on the inclusive cross section
          $\bar{\chi}\chi+(Z\to \ell^+\ell^-)$ at 95\% CLs,
          $\sigma^{95}$ (continuous blue
          line) and $1\sigma$ bands (shaded region), as a function of dark
          sector mass, for negligible splitting and a
          background systematic uncertainty of 1\%.   The expected
          cross section for a pure higgsino with nominal splitting is
          shown for comparison (dashed line).  (Right
          panel) Contours of $\sigma^{95}$ as a function of dark
          sector mass and splitting, for background
          systematic uncertainties of 1-2\%. Results
          were computed for a weak doublet with hypercharge -1/2 and
          nominal splitting, at FCC-hh, for a
          total integrated luminosity of 30 ab$^{-1}$.}
	\label{fig:sig95mupos}
\end{figure}

We plot
in Fig.~\ref{fig:sig95mupos} (left panel) the median expected
exclusion (continuous blue line) as well as the $1\sigma$ bands
(shaded region) as a function of $\chi$ mass, for negligible splitting
and a background systematic uncertainty of 1\%, at
the FCC-hh with an integrated luminosity of 30 ab$^{-1}$.  The LO
cross section for a pure higgsino with nominal splitting is shown as a
dashed blue line, from which we see that the reach in this channel
corresponds to a pure higgsino mass of around 500 GeV.  
We also show in the right-hand panel of
Fig.~\ref{fig:sig95mupos} contours of $\sigma^{95}$ in the
two-dimensional plane $(\mchi,\Delta_+)$, with the contour width reflecting the difference
in the result for 1\% and 2\% systematic uncertainties.  Note that the
limit is rather insensitive to the splitting for small splittings,
but gets weaker for large splittings, as decay products become hard
enough to be visible, and events are vetoed.  The sensitivity in the latter
region could be improved by allowing for additional soft leptons

As mentioned above, our analysis cuts are based purely on the kinematics of t-channel production, rather than details
of the spin or couplings. Hence we expect they can be used as a 
conservative limit on many scenarios featuring pair-produced dark
sector particles of mass $m_\chi$
and small splitting $\Delta_+$, although specific features of other scenarios could of course be exploited to yield tighter constraints.

\subsubsection*{Higher-order corrections}

Sources of sub-leading corrections to our process of interest, and
the results given above 
include:
\begin{itemize}
\item Next-to-leading order (NLO) QCD corrections.  This effect can be
  parameterized by a multiplicative K-factor, $K_{\rm QCD}$, that we
  expect is approximately equal
  for the signal and the irreducible background.  It can yield a 20-30\% boost in the $ZZ$ cross section at
  LHC13 \cite{Mele:1990bq,Neubert:2015fka}, although we expect the effect to be
  relatively smaller at higher scales due to the running of the strong
  gauge coupling.
\item Electroweak sudakov suppression due to large
  logarithms $\mathcal{O}\left(\log{p_T/m_{Z,W}}\right)$ \cite{Becher:2015yea}.  These arise because initial (and typically, final) states are not $SU(2)$ singlets, so
  logarithms do not completely cancel between real and virtual
  correction, and generically give rise to a large suppression of the
  leading order cross section, which increases for increasing $Z$-boson
  $p_T$.  We estimate that the suppression to the diboson
  background is approximately the square of the suppression in the
  signal, leading to no effect on the significance, to first approximation.
\item The effect due to the running of the $SU(2)$ coupling is negligible.
\end{itemize}

Assuming the background is dominated by the irreducible component,
the effect of higher order corrections would be an enhancement in the
overall significance by a factor $\sqrt{K_{\rm QCD}}$, or
equivalently, a reduction in the excluded cross section by the same
factor.  

\section{Pure higgsino thermal relic}
\label{sec:higgsinoRelic}

We will now place these results in a more specific context,
that of a thermal relic in the pure higgsino limit of the MSSM, with
an abundance today that depends on its mass and splittings, the latter
being fixed by mixing with the bino.  We will consider current and
future constraints from direct detection and additional
collider searches that can have sensitivity to the parameter space of
interest.  Indirect detection is not presently sensitive to thermal relic higgsinos in this
mass range \cite{Calibbi:2015nha,Krall:2017xij}, and the large
systematic uncertainties in the background make it difficult to make
precise projections on future sensitivity.  Hence we will neglect it
in the following discussion.

In the limit of small splitting the thermal relic density is
fixed by
the gauge couplings of the higgsino \cite{Giudice:2004tc}:
\be\label{eq:omegaEasy}
\Omega h^2 = 0.105 \Bigl( \frac{\mu}{1~\rm{TeV}}\Bigr)^2 \, .
\ee
Consistency with the current PLANCK measurement of $\Omega h^2 =
0.1198 \pm 0.0026$~\cite{Ade:2015xua} yields an upper limit on the
doublet mass $\mchi$ of [1.05-1.08] TeV.  

\begin{figure}[htb]
\centering
\includegraphics[width=0.8\textwidth]{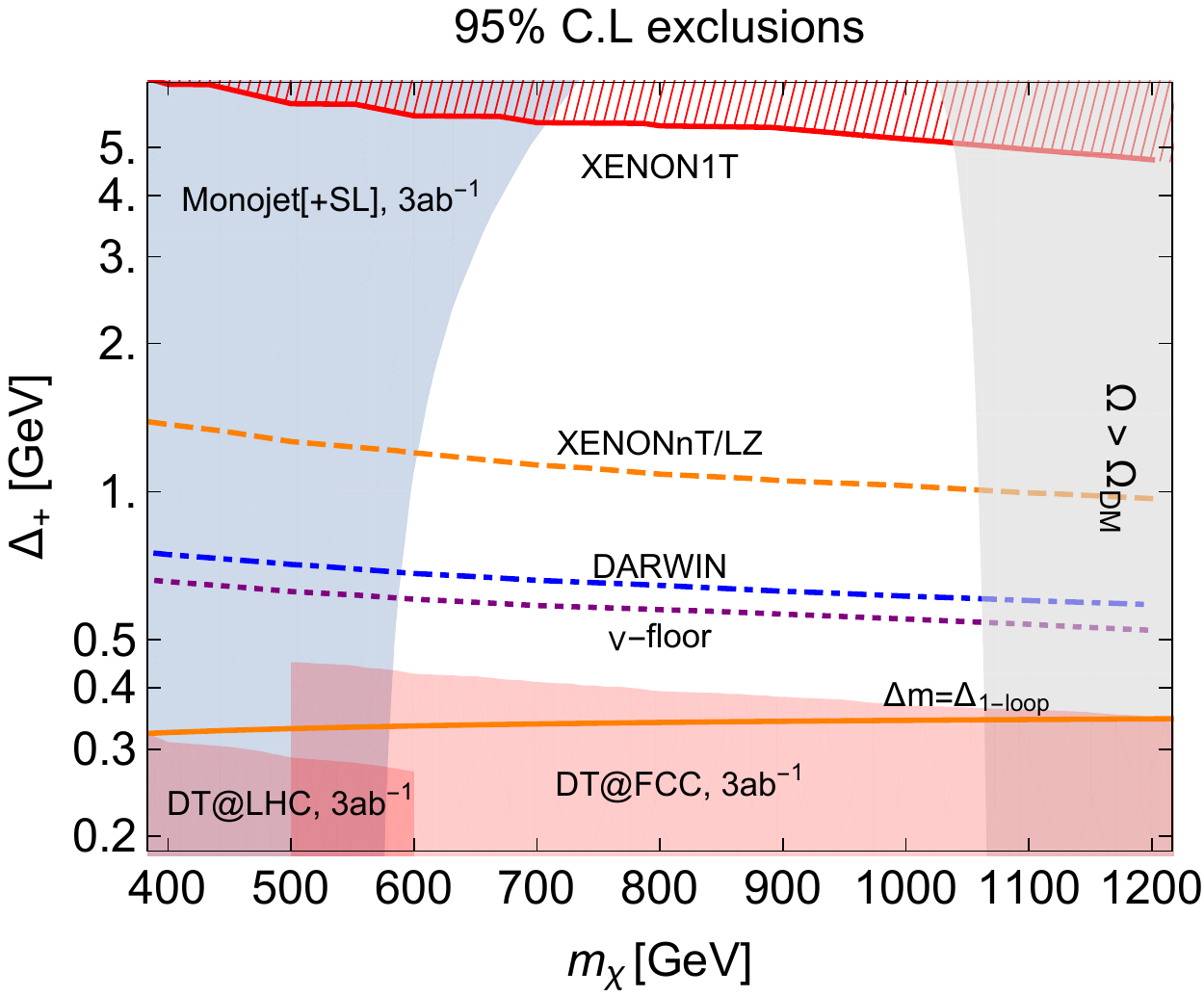}
\caption{Status of the higgsino
  parameter space in view of current and future experimental constraints. We show
  the FCC monojet reach \cite{Low:2014cba} (blue shaded) and the FCC
  disappearing track reach \cite{Mahbubani:2017gjh} (red shaded), for an integrated luminosity at FCC-hh of 3 ab$^{-1}$; as well as
  the region where the thermal relic overcloses the universe (grey shaded). The limit on the spin-independent direct-detection
  cross section due to the recent XENON1T results
  \cite{Aprile:2018dbl} is denoted by a red line, with
  the diagonal-filled region above the line excluded.  The dashed and
  dash-dotted lines correspond to the sensitivity of future direct
 -detection experiments, with the splittings at which the neutrino
 floor becomes relevant shown in purple short dashes. The solid orange line corresponds to the nominal
  1-loop electroweak splitting.}
  \label{fig:MassWindowInitial}
\end{figure}

For larger splittings we
implemented the Lagrangian in Eq.~(\ref{equ:Lagrangian}) in
\texttt{FeynRules v2.3}~\cite{Alloul:2013bka} using the interface to
\texttt{CalcHEP v3.6.27}~\cite{Belyaev:2012qa} in order to employ
\texttt{micrOMEGAs v4.1.2}~\cite{Belanger:2014vza} for the calculation
of dark matter properties.  The regime of higgsino mass and splitting
for which the relic density that would overclose the universe is
shown as a grey shaded region in Fig.~\ref{fig:MassWindowInitial}.

The higgsino parameter space is also constrained by the null results from
direction detection experiments.  In the limit $m_\chi\gg m_S$ (or
equivalently $\mu\gg M_1\gg M_2$ in the MSSM), the relevant interactions are
proportional to the tree-level splittings.
\begin{equation}\label{eq:DD}
L\supset  \frac{g}{2 c_W}\frac{\Delta_+-\Delta_\textrm{1-loop}}{m_Z}\;
h \,\bar{\chi}_1^0 \,\chi_1^0 \;+\; \frac{g}{8c_w}\frac{ \Dzero}{m_\chi}\; \bar{\chi}_1^0 \,\slashed Z\gamma^5\chi_1^0\, ,
\end{equation}
We see from Eq.~(\ref{eq:DD}) that the coupling
for $Z$-boson exchange, which results in a spin-dependent coupling to
matter, is parametrically suppressed, as compared with
the higgs-exchange term, by one to two orders of magnitude for the range of
relic mass relevant to this study.  Moreover the spin structure
functions in spin-dependent detection are small, and don't scale with
detector size due to the difficulty in engineering coherent spin over
macroscopic detector lengths.
Both these effects result in a spin-dependent cross section that is
below the required sensitivity for detection \cite{Calibbi:2015nha}.

We focus instead on the spin-independent cross section with nucleons, which can be
written \cite{Jungman:1995df}: 
\bea\label{eq:ddxsec}
\sigma_{\rm SI/N} &=& g^2_{h\chi\chi} \frac{8 G_F^2 m_Z^2\sin^2{\theta_W}}{\pi}  \left( \frac{m_N}{m_h} \right)^4 \left( \frac{1}{1+\frac{m_N}{\mchi}} \right)^2 \left( \frac{2}{9} + \frac{7 (f_u^N + f_d^N + f_s^N)}{9} \right)^2 
\eea
for nucleon mass $m_N$ and physical Higgs mass $m_h$, where $g_{h \chi
  \chi}$, in the small-mixing limit, is proportional to the tree-level
charged-neutral splitting (see Eq.~(\ref{eq:ddxsec})). Following the recommendation of
  the LHC Dark Matter Working Group \cite{ Boveia:2016mrp} we take equal proton and neutron
  masses, $m_N=0.939$ GeV, and form factors: $f_u^N=0.019$,
  $f_d^N=0.045$~\cite{Hoferichter:2015dsa} and $f_s^N
  =0.043$~\cite{Junnarkar:2013ac}, which gives:
\begin{equation}\label{eq:approxddxsec}
\sigma_{\rm SI/N} \approx  (4 \times10^{-47} \;\;{\rm cm}^2)
\left(\frac{\Delta_+-\Delta_{\rm 1-loop}}{\rm GeV} \right)^2
\end{equation}
for $m_\chi\gg m_N$.  Note that Eq.~(\ref{eq:approxddxsec}) is correct
at first order in the small parameters $m_Z/M_1$, $m_N/\mu$, and is only
dependent on the additional MSSM input parameters $\tan{\beta}$ and
$I=\textrm{Sgn}(\mu)$ indirectly through the value of the tree-level splitting
$(\Delta_+-\Delta_{\rm 1-loop})$ (as well as through terms that are
higher order in $m_Z/M_1$).  We can also see evidence of a `blind
spot' for direct detection \cite{Cheung:2012qy}, where the tree-level splitting goes to
zero for $\tan{\beta}=1$ and $I=-1$.  

We show in 
Fig.~\ref{fig:MassWindowInitial} the region of the compressed higgsino
parameter space that is consistent with the measured thermal relic
density, along with the current exclusion based on the direct
detection constraint recently announced by XENON1T
\cite{Aprile:2018dbl}. We show also projected
sensitivities for future experiments
XENONnT/LZ~\cite{Aprile:2015uzo,Akerib:2015cja} and
DARWIN~\cite{Aalbers:2016jon}; and the neutrino
floor~\cite{Billard:2013qya}, where new techniques will be required to
reject significant backgrounds due to solar neutrinos.  Our results
are computed by numerically diagonalizing the neutralino mass matrix,
and they agree at percent level with
those computed at tree-level using
microOMEGAS.  Loop effects can be consistently included by running and matching the
relevant operators over the different energy scales
\cite{Hill:2013hoa}, this yields an additional contribution to the direct-detection cross section of $\sim 10^{-50}$
cm$^2$ in the pure higgsino limit, which is an order of magnitude
below the level of the neutrino floor.

We add to Fig.~\ref{fig:MassWindowInitial} the collider
constraints acheivable with 3 ab$^{-1}$ of integrated luminosity at
FCC-hh, using a monojet ($+$ soft lepton) search ~\cite{Low:2014cba},
and disappearing charged tracks with improved reconstruction of short
charged tracks ~\cite{Mahbubani:2017gjh}.  We see evidence in this
figure of the true complementarity between the constraints on higgsino
dark matter
relic, and the collider constraints on the dark higgsino.  The measurement of the relic abundance, and indirect
detection and mono-$X$ collider searches, probe the creation and
annihilation of pairs of higgsinos, and are rather insensitive to the
splittings (for small splittings), bounding the higgsino mass from
left and right, respectively.  By contrast, direct detection, which
measures higgsino scattering, and
disappearing charged track searches, which probes chargino decays, are
crucially dependent on the splittings, and constrains these from above
and below.  The combined effect is a `bracketing' of the available
parameter space from all directions.

\begin{figure}[!thb]
	\centering
	\includegraphics[width=0.49\linewidth]{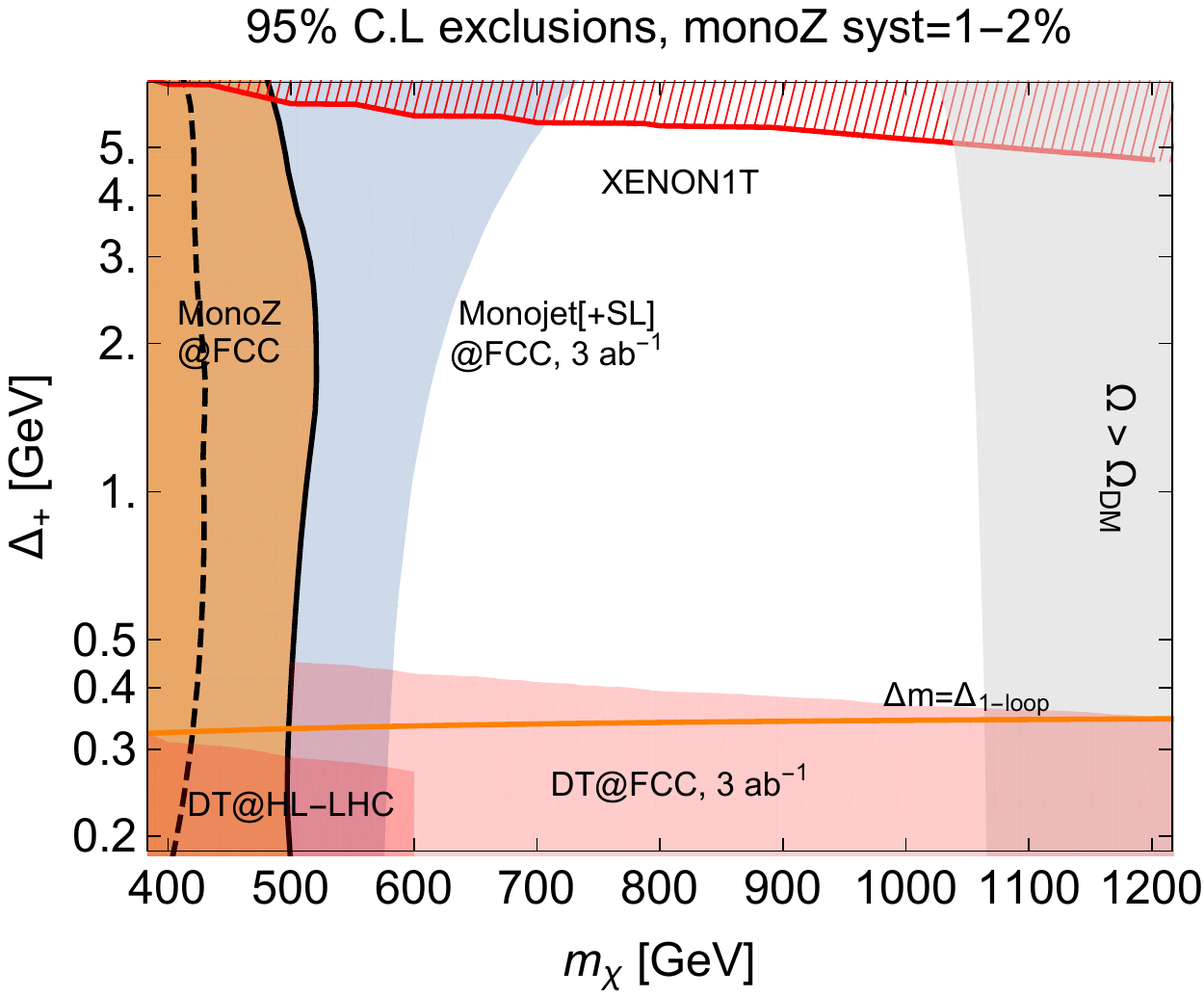}
	\includegraphics[width=0.49\linewidth]{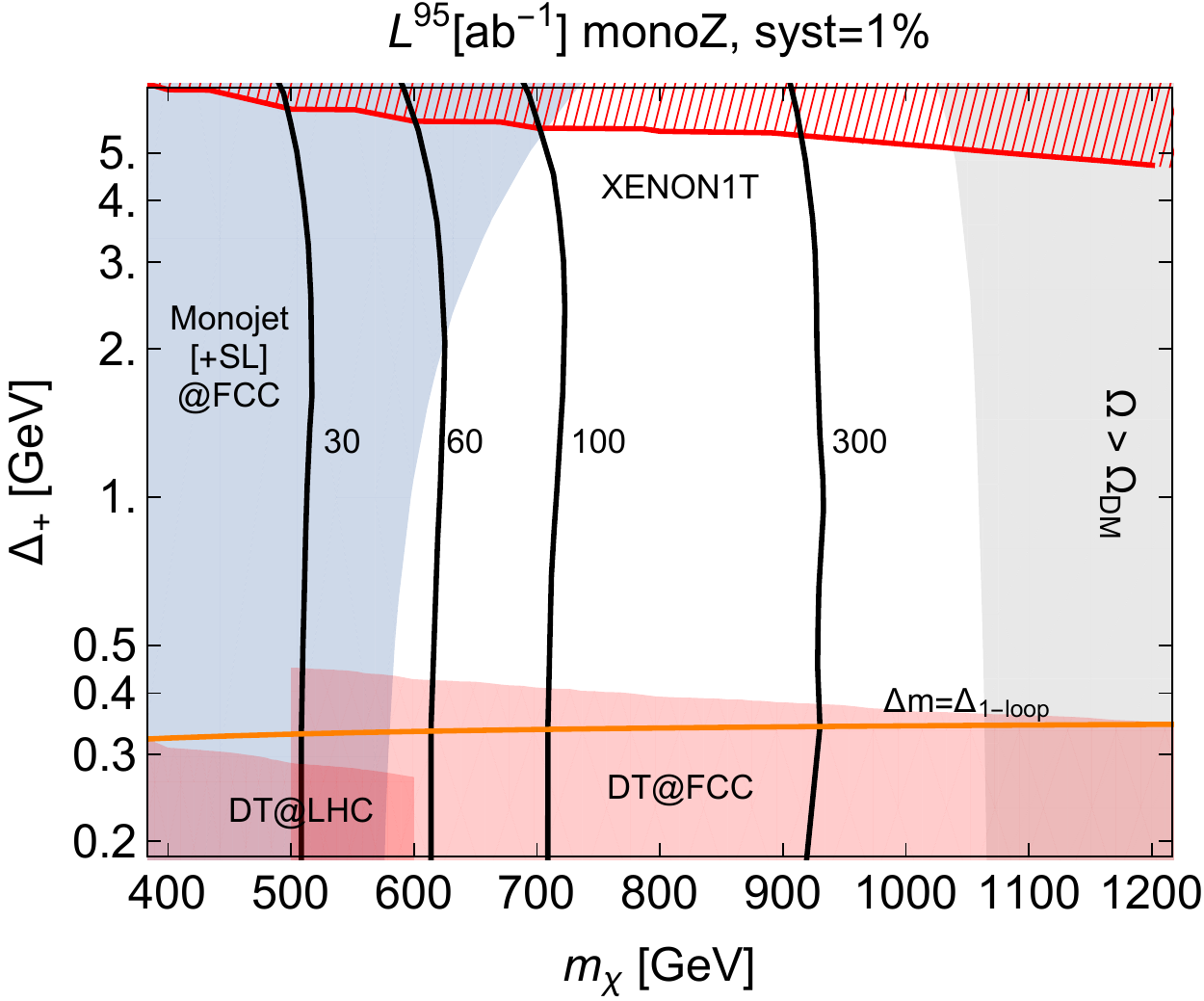}
	\caption{Higgsino parameter space covered by the existing
          studies and the proposed mono-Z search.  (Left panel) The
          95\% CLs exclusion for our dileptonic mono-$Z$ search, for
          30 ab$^{-1}$ of integrated luminosity at FCC-hh, and a
          background systematic uncertainty of 1\% (solid black line)
          and 2\% (dashed black line).  Note that the other collider
          searches shown use a lower luminosity of 3 ab$^{-1}$.  For
          more details see caption of Fig.~\ref{fig:MassWindowInitial}
          and text.  (Right panel)  Contours of luminosity required
          for 95\% CLs exclusion, $L^{95}$, for 1\% systematic
          uncertainty in the background.}
	\label{fig:MassWindowFinal}
\end{figure}

In Fig.~\ref{fig:MassWindowFinal} (left panel) we superimpose on this parameter space
results from our binned mono-$Z$ search with 30 ab$^{-1}$ of
integrated luminosity at FCC-hh and a background systematic
uncertainty of 1\% (solid black line) and 2\% (dashed black line).  We find that the mono-$Z$ search can
probe higgsinos of mass up to 500 (400) GeV, for splittings below 2
GeV, and a background systematic of 1\% (2\%).  For these large
backgrounds the reach is rather
crucially dependent upon how well under control the background
systematics are.  Inclusion of a K-factor for NLO QCD correction, as per the naive
procedure above, would push
the limit up to 550 GeV.  We can recast
this as a limit on winos using the excluded cross section, and obtain
a constraint on the wino mass of 970 GeV.  

In the right panel of Fig.~\ref{fig:MassWindowFinal} we show the
luminosity necessary for a 95\% CLs exclusion on pure higgsinos, for
1\% systematic uncertainty in the background.  Combining 30 ab$^{-1}$
of data from each of two experiments could push the limit above 600 GeV, but closing
the pure higgsino window entirely would require over 600 ab$^{-1}$ of
integrated luminosity, making this goal rather unfeasible in the
dileptonic mono-$Z$ channel.  

\section{Conclusions}
\label{sec:conclu}
In this work we made a careful assessment of the ultimate reach of a
100 TeV pp collider to compressed dark sectors, including some
considerations of higher-order effects, in the dileptonic $Z$+MET
channel.  We arued that dismissing this channel in favour of the
monojet on the basis of a cross-section argument alone is premature.
The backgrounds to mono-$X$ processes are large, making systematic
effects important, and these, particularly for the monojet process,
will likely be more significant in ther busier environment of a 100
TeV proton-proton machine.  

The most promising handle we have on the signal is the MET distribution.  The utility of this variable at
the LHC is rooted in the fact that the pure phase-space behaviour,
which would lead you to expect the two-body SM $ZZ$ distribution to be
harder than that of three-body BSM $\chi\chi Z$ process, is
modified by the presence of a $t$-channel propagator in the production
process, which favours forward $Z$ emission, thus softening the MET
spectrum. This effect is most severe in the
2-body case, and results in increasing signal significance for increasing MET
cut, but only for MET values ranging from the $Z$ mass to the new physics
scale.  Far above the mass of new physics, everything is effectively
massless, and further cuts decrease the sensitivity.

Using a binned CLs method for the total signal and background MET
distributions, we find a 95\% exclusion on pure higgsinos of 500 GeV, with 30
ab$^{-1}$ of integrated luminosity, for a rather optimistic estimate
on the background systematic uncertainty of 1\%.  As with all mono-$X$
searches, this reach is
crucially dependent on the systematics, decreasing to 400
GeV if the systematic uncertainty is doubled.  We find that NLO QCD
effects raise the reach by 50 GeV, while electroweak sudakov
factors have little effect, to first approximation, since their effect
on the background will likely scale like the square of the effect on
the signal.  This reach diminishes for intra-sector splittings of
around 5 GeV; tagging soft decay products could yield additional
sensitivity in this regime.  

Such mono-$X$ searches at future
high-energy hadron colliders would work in concert
with other collider searches, including the
disappearing charged track search, as well as the astro-particle
experimental programme, to bracket the available parameter space
for compressed dark sectors.  However future prospects for discovery
of nearly-pure higgsino dark matter, still look bleak, particularly in
the thermal regime.  It is absurd to imagine that if nature really did work
this way, we may still have no concrete evidence of it half a century from now!
 Although some ideas exist for teasing out the
signature of a pure-higgsino relic from measurements of
neutron stars and compact stars, their sensitivity remains unclear, as
they are subject to large astrophysical uncertainties.  What is clear,
is that this situation is in urgent need of new ideas if we are to
finally close the window on, or discover, electroweakino dark matter.

\acknowledgments 
We would like to thank Pedro Schwaller for collaboration in the early
stages of this work, and Thomas Becher, Benjamin Fuks, Doojin Kim, Gavin Salam, Michele
Selvaggi, Felix Yu and Giulia Zanderighi for useful
discussions. Thanks also to Eilam Gross, Michele Papucci and Josh
Ruderman for sharing their statistics know-how.  We acknowledge the hospitality of the
Galileo Galilei Institute for Theoretical Physics, the Mainz Institute of Theoretical Physics (MITP) and the CERN theory group  during the completion of this work. RM was partially supported by the ERC
grant 614577 ``HICCUP" (High Impact Cross Section Calculations for Ultimate
Precision) and the Swiss National Science Foundation under MHV grant 171330.
\begin{appendix}

\section{Lepton resolution}
\label{sec:leptonres}

\begin{figure}[!tbh]
	\centering
	\includegraphics[width=0.8\linewidth]{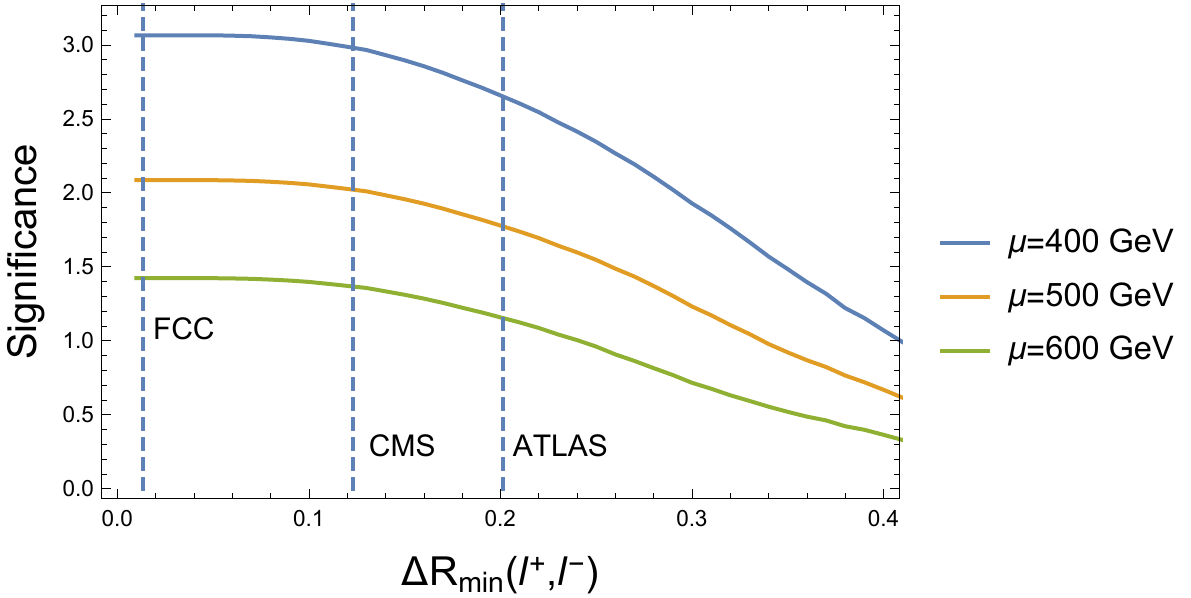}
	\caption{Significances as a function of the minimum resolvable distance between leptons, \drllmin for a nominal $\Dplus=344$ MeV. We have superimposed the \drllmin values corresponding to the \texttt{Delphes} parameterizations of ATLAS, CMS and the FCC detectors. 
	}
	\label{fig:granu}
\end{figure}

As stressed before, this study relies on our capability to reconstruct
a $Z$ from two highly-boosted leptons.  The \texttt{Delphes} FCC card
used when this project started had a \drll  resolution of 0.03. The latest Delphes release further reduces this value to 0.013. It is thus a fair question to understand how the sensitivity degrades with the di-lepton resolution. Thus we present in Fig.~\ref{fig:granu} the significance as a function of \drllmin, where we impose the cut $\Delta R(l^+, l^-) > \Delta R_{\rm min}(l^+, l^-) $, for $\mu$ positive. In addition we superimpose there the  $\Delta R_{\rm min} = (\Delta {\eta}^2 + \Delta {\phi}^2)^{1/2}$ obtained from the \texttt{Delphes} parameterization of the ATLAS, CMS and FCC detectors, where for simplicity we have taken the values corresponding to the central region of the calorimeters, namely $\Delta {\phi} = \pi/18, \pi/36, \pi / 128, \pi / 360$ and $\Delta {\eta}=0.1, 0.087, 0.025, 0.01$ for ATLAS, CMS and the old and latest versions of the FCC respectively.

From the figure we see that even with the current specs for the FCC calorimeter the sensitivity does not degrade significantly, unlike what would happen if the LHC were going to be run with a 100 TeV center-of-mass energy. We thus conclude that resolving these highly-boosted leptons will not be a problem in the future, and we stress once again the importance of considering the detector design to maximize the impact of a future collider for the physics case under consideration.

\section{Profile likelihood}
\label{app:ProfileLikelihood}

We use the profile likelihood method to determine our sensitivity, with
likelihood defined as follows:
\begin{equation}
L(\mu,\bm{\theta})=\prod_{i=1}^{N} \frac{(\mu s_i +
  \theta_i)^{n_i}}{n_i}e^{-(\mu s_i+\theta_i)}\frac{1}{\sqrt{2\pi}\sigma_i}e^{-\frac{(b_i-\theta_i)^2}{2\sigma_i^2}}
\end{equation}
where $\mu$ is the overall normalization of the signal, which has
$s_i$ events in each bin, and $\theta_i$, a nuisance parameter that
will be profiled over, is distributed as a gaussian with mean $b_i$
(the expected number of background events in each bin) and variance $\sigma_i$ given by the systematic
uncertainty.

To find the expected exclusion we compute the usual test
statistic using the Asimov dataset, $\sqrt{q_{\mu,A}}$ (see
e.g. \cite{Cowan:2010js} for details).  This is simple enough to do
analytically, and yields:
\begin{equation}
q_{\mu,A}=\sum_i\left[-2b_i\log{\left(\frac{\mu
        s_i+\hat{\theta}_A}{b_i}\right)}  -2(b_i-\mu s_i-\hat{\theta}_A)-\frac{(b_i-\hat{\theta}_A)^2}{\sigma_i^2}\right]
\end{equation}
for
\begin{equation}
\hat{\theta}_A=\frac{1}{2}\left[b_i-\mu
  s_i-\sigma_i^2+\sqrt{\left(b_i+\mu
      s_i-\sigma_i^2\right)+4b_i\sigma_i^2}\right]\; .
\end{equation}
In the limit $s_i\ll b_i+\sigma_i^2$ this reduces to the sum over the
familiar rule-of-thumb expression $s_i/\sqrt{b_i+\sigma_i^2}$ for each bin.  The CLs
p-value is given in the asymptotic limit by \cite{Cowan:2010js}:
\begin{equation}
p_{\mu,A}=\frac{p_s}{1-p_b}=\frac{1-\Phi(\sqrt{q_{\mu,A}})}{\Phi(0)}
\end{equation}
and we can compute the exclusion at 95\% confidence by numerically solving for
the normalization factor $\mu^{95}$ that gives $p_{\mu,A}=0.05$, at
each signal mass $m_\chi$.

\end{appendix}

\bibliographystyle{JHEP}
\bibliography{MonoZ}

\end{document}

%% file: tables/cutflow.tex
\begin{tabular}{ccccccc}
	\noalign{\hrule height 1pt}
	Process & Preselection & $x< 1.3$ & $y < 1.4$ & $\metm > 550$
                                                        GeV & $\metm > 900$ GeV    \\
	\noalign{\hrule height 1pt}
	Signal A & 176 & 168 & 163  &  & 72    \\
	Signal B & 1234 & 1184 & 1166 & 914 &   &  \\
	\noalign{\hrule height 1pt}
	$ZZ \to l^+ l^- \nu \nu$            & 83799   & 81292    &
                                                                   81178   & 51623 & 11172  \\
	$W^+ W^- \to l^+ \nu l^- \nu$       & $< 1$   & $< 1$    & $<1$
                                                      & $< 1$ & $< 1$ \\
	$tt \to l^+ b \nu l^- \bar{b} \nu $ &  5136   & 4484     &3980
                                                      &2300 & 238\\
	$tt \to l \nu b \bar{b} j j $       & 57161   & 44459    & 5835
                                                      & 3922 & 635   \\
	$(Z \to l^+ l^-)$ +jets             & 321204 &  208220  &
                                                                  48719  & 25174 & 4373  \\
	$(W \to l \nu)$+jets                & 2142    & 1797     & 72
                                                      & 66 & 24  \\
	$ZW \to l^+ l^- l \nu$              & 11835   & 10891    &
                                                                   10873
                                                      & 5935 & 889  \\
	$ZZ \to l^+ l^- j j$                & 2021     & 819      &
                                                                    339
                                                      & 137 & 13  \\  
	$ZW \to l^+ l^- j j$                & 301     &  113   &  36
                                                      & 15 & 3 \\
	\noalign{\hrule height 1pt}
	$10^3$ S$_{\rm A}$/B           & 0.36 & 0.48   & 1.08   & &  4.14  \\
	Significance A ($\beta = 0 $) &0.25    & 0.28  & 0.42 & & 0.55  \\
	Significance A ($\beta = 0.01 $)& 0.04 & 0.05 & 0.10  & & 0.33  \\
	\noalign{\hrule height 1pt}
	$10^3$ S$_{\rm B}$/B            & 2.55 & 3.36  & 7.72 & 10.26 &  & \\
	Significance B ($\beta = 0 $) & 1.77 & 1.99  & 3.00  & 3.06  & &\\
	Significance B ($\beta = 0.01 $)& 0.25 & 0.33  & 0.75  & 0.97 & &\\	
		\noalign{\hrule height 1pt}
	\noalign{\hrule height 1pt}

\end{tabular}
